\documentclass{mn2e}
\usepackage{psfig}

\newif\ifAMStwofonts
\AMStwofontstrue

\newcommand{\ltsimeq}{\raisebox{-0.6ex}{$\,\stackrel
{\raisebox{-.2ex}{$\textstyle <$}}{\sim}\,$}}
\newcommand{\gtsimeq}{\raisebox{-0.6ex}{$\,\stackrel
{\raisebox{-.2ex}{$\textstyle >$}}{\sim}\,$}}

\title[Polarimetry of HL Tau] {High resolution imaging polarimetry of HL Tau and magnetic field 
structure$^{*}$}
\author[Lucas, Fukagawa et al.]
{P.W.Lucas$^{1}$, Misato Fukagawa$^{2}$, Motohide Tamura$^{3}$, A.F.Beckford$^{1}$,
\newauthor Yoichi Itoh$^{4}$, Koji Murakawa$^{5}$, Hiroshi Suto$^{5}$, Saeko S. Hayashi$^{5}$,
\newauthor Yumiko Oasa$^{4}$, Takahiro Naoi$^{6}$, Yoshiyuki Doi$^{5}$, Noboru Ebizuka$^{7}$,
\newauthor and Norio Kaifu$^{3}$\\
$^{*}$ Based in part on data collected at Subaru Telescope, which is 
operated by the National Astronomical Observatory of Japan.\\
$^{1}$Dept. of Physical Sciences, University of Hertfordshire, College Lane,
Hatfield AL10 9AB, UK\\ email: pwl@star.herts.ac.uk\\
$^2$Dept. of Astronomy, University of Tokyo, 7-3-1 Hongo, Bunkyo, Tokyo
113-0033, Japan  \\
$^3$National Astronomical Observatory of Japan, 2-21-1, Osawa, Mitaka, Tokyo
181-8588, Japan \\
$^4$Graduate School of Science and Technology, Kobe University, 1-1 Rokkodai,
Nada, Kobe 657-8501, Japan \\
$^5$Subaru Telescope, 650 North A'ohoku Place, Hilo, HI 96720, USA\\
$^6$Dept. of Earth and Planetary Science, University of Tokyo, 7-3-1 Hongo,
Bunkyo, Tokyo 113-0033, Japan  \\
$^7$The Institute of Physical and Chemical Research, 2-21 Hirosawa, Wako,
Saitama 351-0198, Japan}

\date{Accepted - . Received 2003 November}
\pagerange{\pageref{firstpage}--\pageref{lastpage}}
\pubyear{2003}

\begin{document}
\label{firstpage}
\maketitle

\begin{abstract}
We present high quality near infrared imaging polarimetry of HL Tau at 0.4 to 0.6 arcsec 
resolution, obtained with Subaru/CIAO and UKIRT/IRCAM. 3-D Monte Carlo modelling
with aligned oblate grains is used to probe the structure of the circumstellar 
envelope and the magnetic field, as well as the dust properties. 
At J band the source shows a centrosymmetric pattern 
dominated by scattered light. In the H and K bands the central source 
becomes visible and its polarisation appears to be dominated by dichroic 
extinction, with a position angle inclined by $\approx 40^{\circ}$ to the disc axis.
The polarisation pattern of the environs on scales up to 200 AU is consistent with 
the same dichroic extinction signature superimposed on the centrosymmetric scattering pattern.
These data can be modelled with a magnetic field which is twisted on 
scales from tens to hundreds of AU, or alternatively by a field which is globally 
misaligned with the disc axis. A unique solution to the field 
structure will require spatially resolved circular polarisation data.
The best fit Monte Carlo model indicates a shallow near infrared extinction law. 
When combined with the observed high polarisation and non-negligible albedo
these constraints can be fitted with a grain model involving dirty water ice 
mantles in which the largest particles have radii slightly in excess of 1~$\mu$m. 
The best fit model has an envelope structure which is 
slightly flattened on scales up to several hundred AU. Both lobes of the bipolar 
outflow cavity contain a substantial optical depth of dust (not just within the
cavity walls). Curved, approximately parabolic, cavity walls fit the data better than
a conical cavity. The small inner accretion disc observed at millimetre wavelengths is not 
seen at this spatial resolution. 
\end{abstract}


\section{Introduction}

  HL Tau is among the most well studied of all Young Stellar Objects (YSOs).
It is a low mass YSO in the nearby Taurus-Auriga star formation region with 
a spectral energy distribution which is Class I but relatively flat between 
wavelengths 2 to 60~$\mu$m (Men'shchikov, Henning \& Fischer 1999). This indicates an 
evolutionary status fairly close to the boundary between Class I and Class II 
YSOs, sometimes known as flat spectrum T Tauri stars (Adams, Lada \& Shu 1988).
Such sources are astrophysically very useful. They often retain a large 
circumstellar envelope, which can provide clues to the formation process,
but the optical depth of the envelope is low enough that the central regions
can be observed in the near infrared waveband with high spatial resolution and
high signal to noise ratio. The system is surrounded by an unusually massive
envelope ($\sim 0.1$~M$_{\odot}$, Beckwith et al. 1986; 1990) with a radius of 
approximately 1300 AU. It has been claimed that this structure is rotating 
(Sargent \& Beckwith 1991) or infalling (Hayashi et al. 1993) but later 
work (Cabrit et al. 1996) casts doubt on these suggestions and indicates that the 
kinematics are complicated by the entrained outflow. Recent $^{13}$CO mapping of 
the large scale environment (Welch et al. 2000) finds that the kinematics are further
confused by the presence of an expanding shell around the adjacent T Tauri star 
XZ Tau, which has recently reached the location of HL Tau.

  Close et al.(1997) imaged HL Tau at 0.2 to 0.3 arcsec resolution with adaptive 
optics (AO). They observed a compact irregular nebula containing a point source 
which was bright in the H (1.65~$\mu$m) and K (2.2~$\mu$m) bands but less
prominent in the J band (1.25~$\mu$m) where the extinction is higher.
Optical imaging at even higher spatial resolution with HST (Stapelfeldt et al.
1995) showed only circumstellar nebulosity and no point source, demonstrating
the embedded nature of the system.

  Imaging provides only limited information on the structure of a YSO nebula
the dust grains within it. This is particularly true in systems like HL Tau
where the central source is so bright at 2~$\mu$m that details of the 
circumstellar matter are lost in the skirts of the image profile. Imaging 
polarimetry circumvents this latter problem (the central source being much 
fainter in polarised light) and provides a wealth of additional information
on the nature of circumstellar dust grains and the physical structure of the 
envelope (eg. Tamura et al.1991; Whitney \& Hartmann 1993; Fischer, Henning \& 
Yorke 1994; 1996; Gledhill et al. 1996; Lucas \& Roche 1997, 1998 (hereafter LR97, 
LR98); Whitney et al. 1997). Polarimetry can also yield information on the magnetic 
field structure within the circumstellar envelope, since non-spherical dust 
grains in the protostellar environment are believed to rapidly align their axis of 
greatest rotational inertia so that it precesses around the axis of the local 
magnetic field. Note that the degree of grain alignment and the dominant
alignment mechanism remain uncertain (radiative torque alignment or paramagnetic 
alignment aided by H$_{2}$ rotation torques; see recent review by Lazarian 2003)
but it is reasonably certain that the sense of the grain precession axis is parallel 
to the magnetic field direction (Draine \& Weingartner 1997).

  Near infrared polarimetry of point source Class II and Class III YSOs usually 
reveals only a low degree of polarisation due to dichroic extinction by 
aligned grains. Polarimetry of Class I YSOs with circumstellar nebulae often 
shows a region of aligned vectors known as a 'polarisation disc' at the 
location of the central source, and a gradual transition to a centrosymmetric 
pattern of vectors in the surrounding nebula. These polarisation discs are 
often attributed to multiple scattering in cases where the optical depth 
toward the central source is too high for direct observation (Whitney et 
al.1993; Bastien \& Menard 1988). However, polarisation discs are also 
sometimes observed in cases where the optical depth is low (eg. TMR-1, 
see LR97) and the multiple scattering mechanism does not easily reproduce the 
high polarisation ($>15\%$) of the polarisation disc seen toward some sources 
(eg. IRAS 04302+2247, see LR97). Alternative mechanisms must therefore be
considered also, including: (1) scattering and extinction by aligned 
non-spherical grains (Gledhill \& McCall 2000; Whitney \& Wolff 2002; 
Lucas 2003; Lucas et al. 2003); and (2) artifacts of the seeing profile, which 
we call the 'illusory disc' (see Whitney et al.1997 and LR98).

  HL Tau was therefore identified as an ideal subject for imaging polarimetry
at high spatial resolution in the JHK bands, since at the 0.4 arcsec resolution
commonly achieved in good seeing conditions at Mauna Kea Observatory (without Adaptive 
Optics) it was anticipated that the central source would be obscured by 
nebulosity at J band, but become visible in the H and K bands. This proved to 
be the case, as previously found by Beckwith \& Birk (1995) in 1 arcsec resolution
data. The data collection is described in section 2, the observational 
results are described in section 3, and sections 4 and 5 detail the methodology 
and results of 3-D Monte Carlo modelling of the system with aigned grains. 
The conclusions are listed in section 6.

\section{Observations}

  Near infrared imaging polarimetry was independently carried out with 2 telescopes 
at Mauna Kea Observatory on dates a few weeks apart, providing a useful test
of the reliability of the data. The 3.8-m United Kingdom Infrared Telescope 
(UKIRT) was used on 28th December 2000 by observatory staff as part of the UKIRT
Service programme, using flexible scheduling to ensure good seeing conditions. 
Observations were made in the J, H and K bands with the camera IRCAM. 
The 8.2-m Subaru Telescope was used on 17th January 2001 during commissioning of the 
Coronographic Imager with Adaptive Optics (CIAO, see Tamura et al. 2000). The Subaru 
observations were conducted in H and K bands, but the high quality data was obtained 
only in the K band. The H band data was taken at high airmass ($\gtsimeq$ 1.8) and 
with less integration time and suffered from imperfect tracking which obscured small 
scale structures. We will not discuss the Subaru H band data quantitatively.
Although the Subaru AO system was not used, good seeing conditions 
allowed 0.4 arcsec resolution to be obtained (full-width half maximum). 
Both datasets were taken in photometric conditions. The UKIRT dataset suffered from an 
offset from correct focus which slightly reduced the spatial resolution and data quality 
(see below). However the spatial resolution was still 0.4 to 0.6 arcsec and the focus tracking 
system provided excellent image stability, which is indicated by both the consistency of 
subsets of the polarisation measurements in small apertures (see Section 3.2) and the 
good agreement between the Subaru and UKIRT datasets.

\subsection{Subaru data}

  The Subaru/CIAO observations were conducted with a single beam system, using
a cold wire grid analyser. A half wave plate modulator is placed upstream of 
any optical components except telescope optics including adaptive optics 
system, which minimizes the effect of AO mirrors on instrumental polarisation (IP), 
resulting in IP$<1\%$. The image scale on the 1024$^2$ InSb ALADDIN 
\newcounter{two}\setcounter{two}{2}\Roman{two} array was 0.022 arcsec 
pixel$^{-1}$. The warm modulator was rotated in sequence through position angles 
0$^{\circ}$, 22$^{\circ}$.5, 45$^{\circ}$, 67$^{\circ}$.5 to measure the 
Stokes I, Q and U parameters. HL Tau was observed with 
read noise limited integrations of 1~s, and 10 exposures were coadded into one image frame. 
Before rotating the modulator, 6 image frames were taken (i.e. 60~s of data). The 
integration time per modulator cycle was 240~s 
(4 modulator positions) and 4 consecutive modulator cycles were performed at
different dither positions on the array. The total on source integration time at 
K band was 960~s. 
A polarised star HD251204 (Weitenbeck 1999) was observed following HL Tau, 
providing a reliable determination of the position angle calibration. 
Several other standards taken at the later commissionings of the CIAO 
polarimeter gave consistent values for the position angle calibration. 
The polarisation efficiency was measured with an unpolarized star and 
another wire grid analyser inserted over the half wave plate, 
making 100\% polarisation artificially. The efficiency of 97\% in K band 
was applied for the calibration of observed polarisation degree. 
Individual frames were calibrated in the usual manner with the reduction 
packages in IRAF: dark subtraction, flat-fielding by dome-flats, bad-pixel 
substitution, and sky subtraction. Since there is no difference between 
flats taken at four modulator angles, all flats were combined and used 
in order to obtain higher signal-to-noise. The dithered frames taken 
just before and after the calibrated frame were combined to use as a sky frame
for background subtraction.\\

\subsection{UKIRT data}

  The UKIRT/IRCAM observations were conducted with a dual beam system with a 
cold Wollaston prism inside IRCAM and a half waveplate modulator mounted in 
IRPOL-2 (see Hough et al., 1995). Dual beam systems reduce the effect of variations 
in seeing and transparency between different rotation settings of the modulator.
The warm modulator was rotated in the same sequence as Subaru.
The image scale on the 256$^2$ InSb array was 0.08 arcsec pixel$^{-1}$, following 
the reconfiguration of the camera as IRCAM-TUFTI in early 1999. A warm mask 
mounted on the camera entrance window prevents overlap of the dual beams, and
divides the useful observing area on the sky into 2 parallel strips 
20 arcsec in length and 4 arcsec wide. HL Tau was observed in 10~s observation blocks
at each waveplate position before rotating the modulator or dithering the 
target position, in order to smooth out the effect of any short term variations in 
the image profile. The total coadded exposure time was 240~s at J band, 
240~s at H band and 360~s at K band but individual exposures were short
and read noise limited, due to the brightness of the target.
No polarisation standards were observed with UKIRT but observatory staff have 
found the IRPOL-2 system to be stable over many years, with polarisation 
efficiency $>99\%$ and a consistent position angle calibration relative to north
($6.7 \pm 0.1^\circ$). Note that IRPOL2 lies above IRCAM in the light path, so the camera 
affects the measured polarisaton only through (i) the efficiency of the Wollaston in IRCAM
($>99\%$); and (ii) a rotation of the image plane with respect to north which must be
reversed in order to match the image data to the polarisation data. The image 
rotation is $-88.84 \pm 0.08^\circ$, see www.jach.hawaii.edu/JACpublic/UKIRT/home.html.
The UKIRT data were dark subtracted and flatfielded in IRAF before commencing
polarimetric reduction using the Starlink package POLPACK. Individual frames were
registered using the central flux peak of HL Tau, which is very well defined in the H and
K bands, permitting precision of a small fraction of a pixel. The J band registration 
using the more nebulous flux peak was less precise but still accurate to 1 pixel or 
better (0.08 arcsec).\\
\vspace{-5cm}

\psfig{file=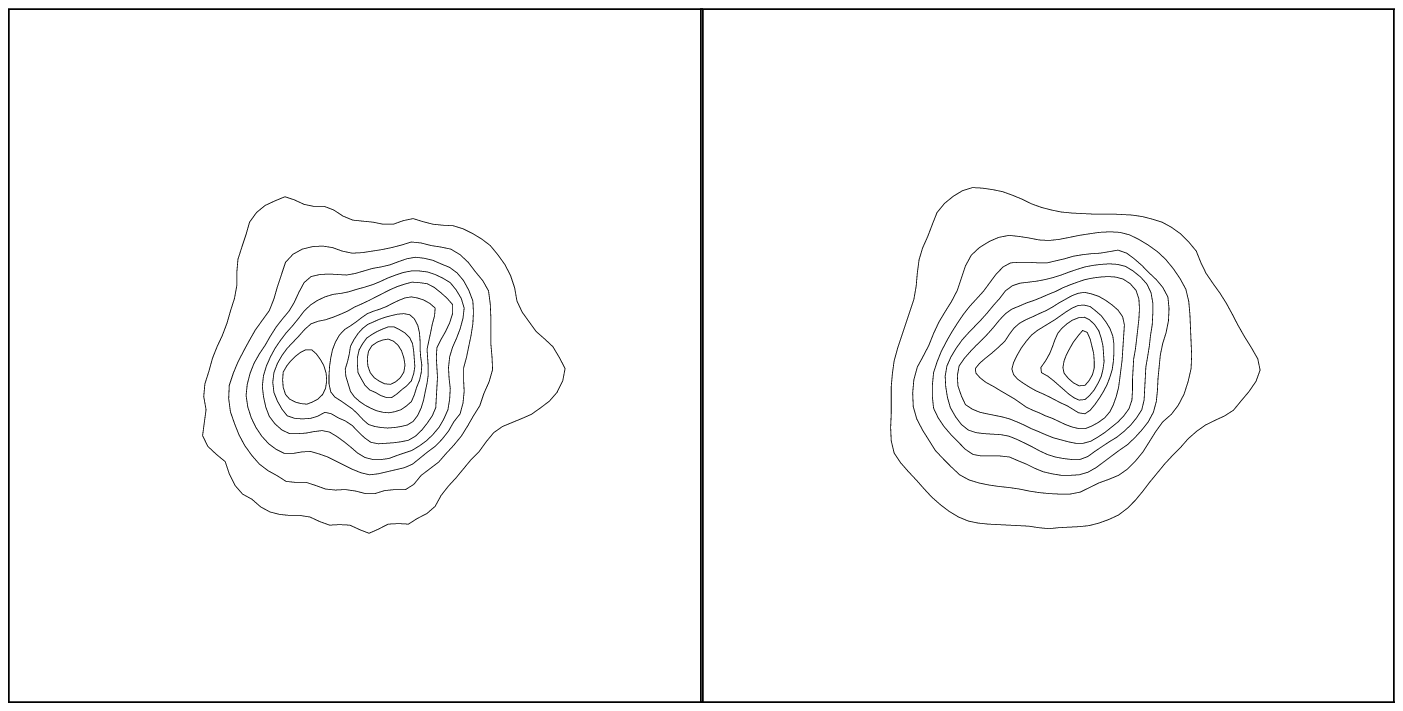,clip=,width=10cm}
\vspace{-4cm}
Figure 1. UKIRT image profile from 27 December 2000. (left) as observed.
(right) after rebinning to the 0.15 arcsec pixel scale used in the models in Section 5.
Contour levels are 0.1, 0.2, 0.3 ... 0.9, normalised to unity at the peak. \\
\\




{\bf Focussing Problem}\\

The focus offset mentioned above caused the UKIRT K band data to show a small secondary peak
on the shoulder of the image profile (see Figure 1) located 0.4 arcsec from the principal peak at 
P.A.=103$^{\circ}$ (east of north) and with $\sim 90\%$ of the peak intensity.
Inspection of K band archival imaging data of an 
isolated star taken on the previous night (27 December) clearly shows the secondary source 
appearing at the same location as the focus was changed in a sequence from in focus to far out of 
focus. This introduced the spurious secondary source in the HL Tau K band data, which shows the same 
percentage and position angle of polarisation as the main peak. The secondary source had $\sim 70\%$ 
of the peak intensity in the profile of the isolated star but appears at a higher level in the HL 
Tau data due to the bright underlying nebulosity. Additional archival imaging polarimetry 
taken 2 weeks later in very good seeing conditions (0.3 to 0.4 arcsec) shows that there is no sign 
of this secondary source when the telescope is well focussed, confirming that the artefact was not 
caused by the Wollaston prism or the IRPOL-2 waveplates.

Despite the focus offset the UKIRT tip/tilt secondary mirror and 
focus tracking system ensured that the image profile was very stable, with no variation 
being detectable between the 36 individual K band frames or the J and H band frames. This is 
confirmed by the consistent results of small aperture polarimetry in independent cycles of the 
modulator (see section 3.2). The effect of the problem is therefore simply to reduce the spatial
resolution of the data slightly to approximately 0.55 arcsec (see end of section 4).\\

{\bf 3-micron Spectrum}\\

  An L band spectrum of HL Tau was taken with UKIRT and the near infrared
spectrograph CGS4 on 1st December 1998. These data were also taken in Service mode
by observatory staff. The data were obtained in order to search for the 'Unidentified
Infrared Bands' sometimes seen in more massive YSOs, which are often attributed
(Sellgren 1984) to UV excited emission by ionised polycyclic aromatic hydocarbons (PAHs).
A 40 l/mm grating was used, giving a spectral resolution of R=1320 and a wavelength span
of 3.0 to 3.62$~\mu$m. While no PAH emission was detected, the strong water ice 
absorption seen in the spectrum is relevant to the discussion of grain composition in
Section 5.2.3.


\section{Results}


\begin{figure*}
\begin{center}
\begin{picture}(200,480)

\put(0,0){\includegraphics{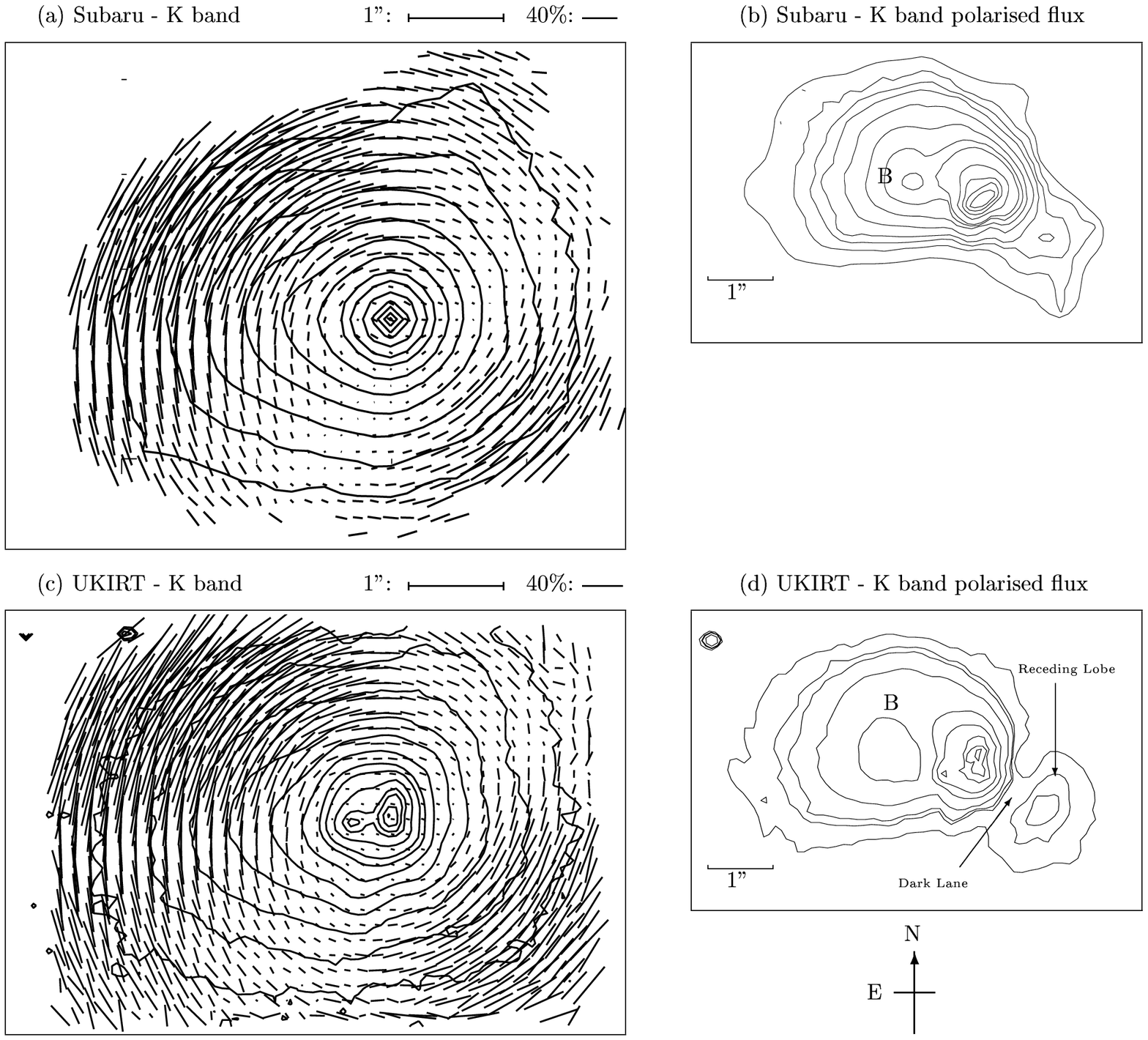}}

\end{picture} 
\end{center}
Figure 2. K band data. (a) Subaru image and polarsation map; (b) Subaru polarised
flux image; (c) UKIRT image and polarisation map; (d) UKIRT polarised flux map.
The data from both telescopes are very similar. The receding lobe of the bipolar
nebula appears at lower contrast in the Subaru polarised flux data than the UKIRT data
because the central peak is more highly polarised and better resolved in the Subaru data (see
Table 1). The polarisation vectors are 8$\times$8 binned in (a) and 2$\times$2 binned in (c), 
giving pixel scales of 0.17 arcsec and 0.16 arcsec for the Subaru and UKIRT data 
respectively. Contour levels are normalised to unity at the peak. The levels
in (a) and (c) are: 0.0031, 0.0063, 0.0125, 0.025, 0.05, 0.1, 0.2, 0.3, 0.5, 0.7, 0.8, 0.9, 0.99.
Contour levels in (b) are 0.028, 0.05, 0.061, 0.1, 0.2, 0.3, 0.5, 0.7, 0.9. Levels 
in (d) are 0.019, 0.029, 0.039, 0.049, 0.07, 0.1, 0.137, 0.174, 0.3, 0.5, 0.7, 0.9, with
the lowest contour level averaged using 2$\times$2 pixel binning.

\end{figure*}

Polarisation maps and polarised flux images in the K, H and J bands are shown
in Figures 2(a-d), 3(a-c) and 4(a-b). More detailed K and H band polarisation maps of the central
regions are shown in Figure 5(a-c). The structure of the HL Tau nebula is indicated
by contours overlaid on the polarisation maps. The flux distribution is dominated by
a point source in the K and H bands but this suffers sufficient 
extinction within the system that circumstellar nebulosity is easily visible at the 0.4 
arcsec resolution of the data. At J band the optical depth is sufficient to obscure the 
point source at this spatial resolution. The J band flux peak is dominated by scattered light
and is offset from the centre of the polarisation pattern by $0.40 \pm 0.11$ arcsec along the disc 
axis. (We define the centre of the pattern using the polarisation nulls seen to either side of the 
region of low polarisation located southwest of the flux peak in Figure 4(a).) 
This contrasts with the observations of Close et 
al.(1997), wherein the higher spatial resolution (0.3 arcsec) reduced the nebulous flux in the beam 
by a factor of $\sim 2$ relative to these data, so that the point source was marginally detectable.
Close et al. subracted a model of the nebulosity based on Hubble Space Telescope I band
data (Stapelfeldt et al. 1995) to reveal the point source more clearly.
The polarisation of the J band flux peak in our data is rotated by $\sim 50^{\circ}$ compared
to the position angle seen at H and K and has an approximately centrosymmetric orientation 
(see section 3.2), confirming that we observe a scattering peak at J band. 

The UKIRT K band data show the small secondary flux peak 0.4 arcsec east of the central point source 
which is the image artefact caused by the focussing offset described in Section 2.2. The H and J band 
data also show an easterly elongation of the contours within 0.4 arcsec of the surface brightness peaks 
in Figures 3(a) and 4(a) but the false secondary peak is not resolved since the surface distributions 
are too smooth at these wavelengths. 

The central point source is located at the approximate focus of the centrosymmetric pattern of
polarisation vectors in the surrounding nebula. In the Close et al.(1997) data its spatial location 
is the same at J, H and K to within 0.1 arcsec so it is therefore simply interpreted as the 
protostar itself. In later sections of this paper the term 'protostar' is used synonymously with the 
central point source which is observed here only at H and K.
Most (but not all) other researchers have made the same interpretation after observing the near
concidence of both the millimetre source and the centimetre source with the infrared source (see 
Mundy et al.1996; Wilner, Ho \& Rodriguez 1996; Rodriguez et al.1992; 1994). This is further supported 
by the presence of the optical jet extending away from this location (Mundt et al. 1990; Bacciotti, 
Eisloffel \& Ray 1999).

Lower resolution infrared polarisation maps of the object have been made by 
Yamashita, Hodapp \& Moore (1994); Weintraub, Kastner \& Whitney (1995); Takami et al.(1999); 
Bastien et al.(private comm.). LR98 published Shift and Add polarimetry of HL Tau in the H and L bands
at similar spatial resolution to these data but with much lower signal to noise.
The optical polarisation structure has been mapped by Gledhill \& Scarrott (1989).
These previous lower resolution polarisation maps show the same centrosymmetric structure as 
observed here but did not resolve the distinct position angle of polarisation of the central point 
source, seen in the UKIRT and Subaru data (see section 3.3). The dense accretion disc 
indicated by the dark lane seen in polarised flux in Figure 2(d) was 
also observed with the same orientation by Yamashita et al.(1994). Weintraub et al.(1995) and LR98
also deduced this disc orientation from the region of low polarisation seen in the polarisation
maps between the two highly polarised outflow lobes. 

\begin{figure*}
\begin{center}
\begin{picture}(200,650)

\put(0,0){\includegraphics{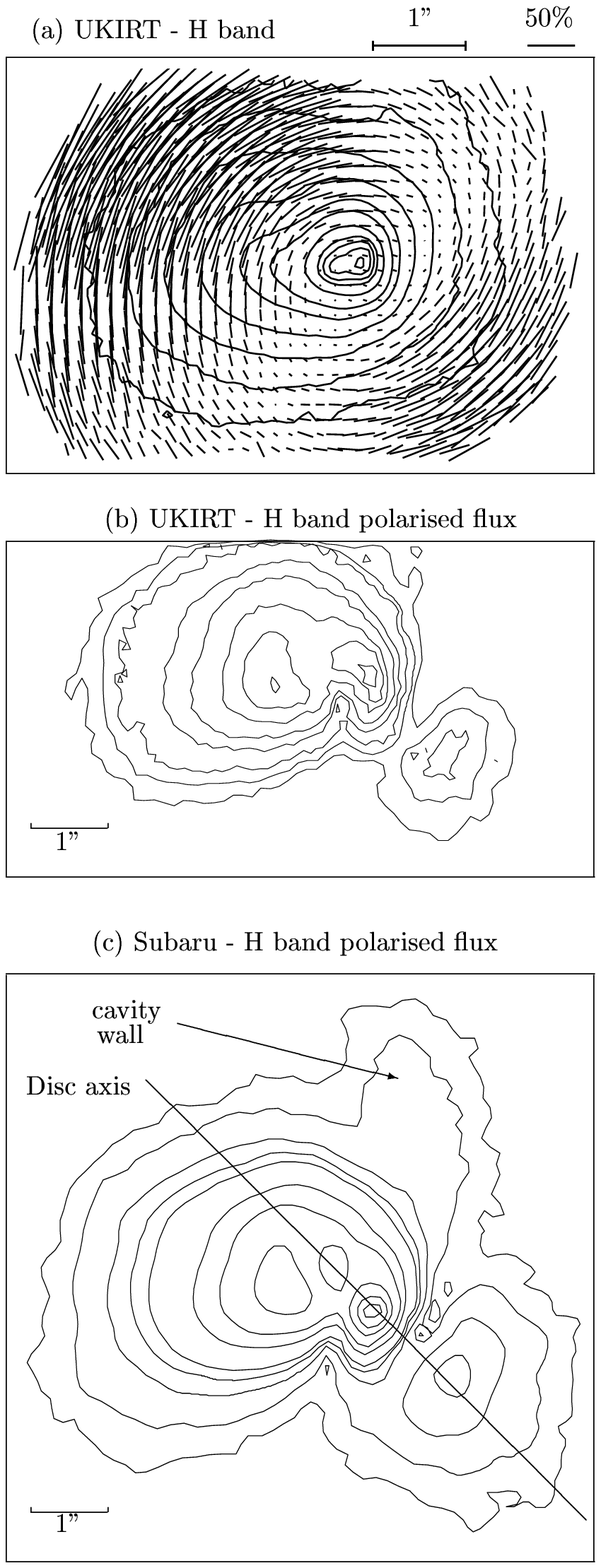}}

\put(0,0){\includegraphics{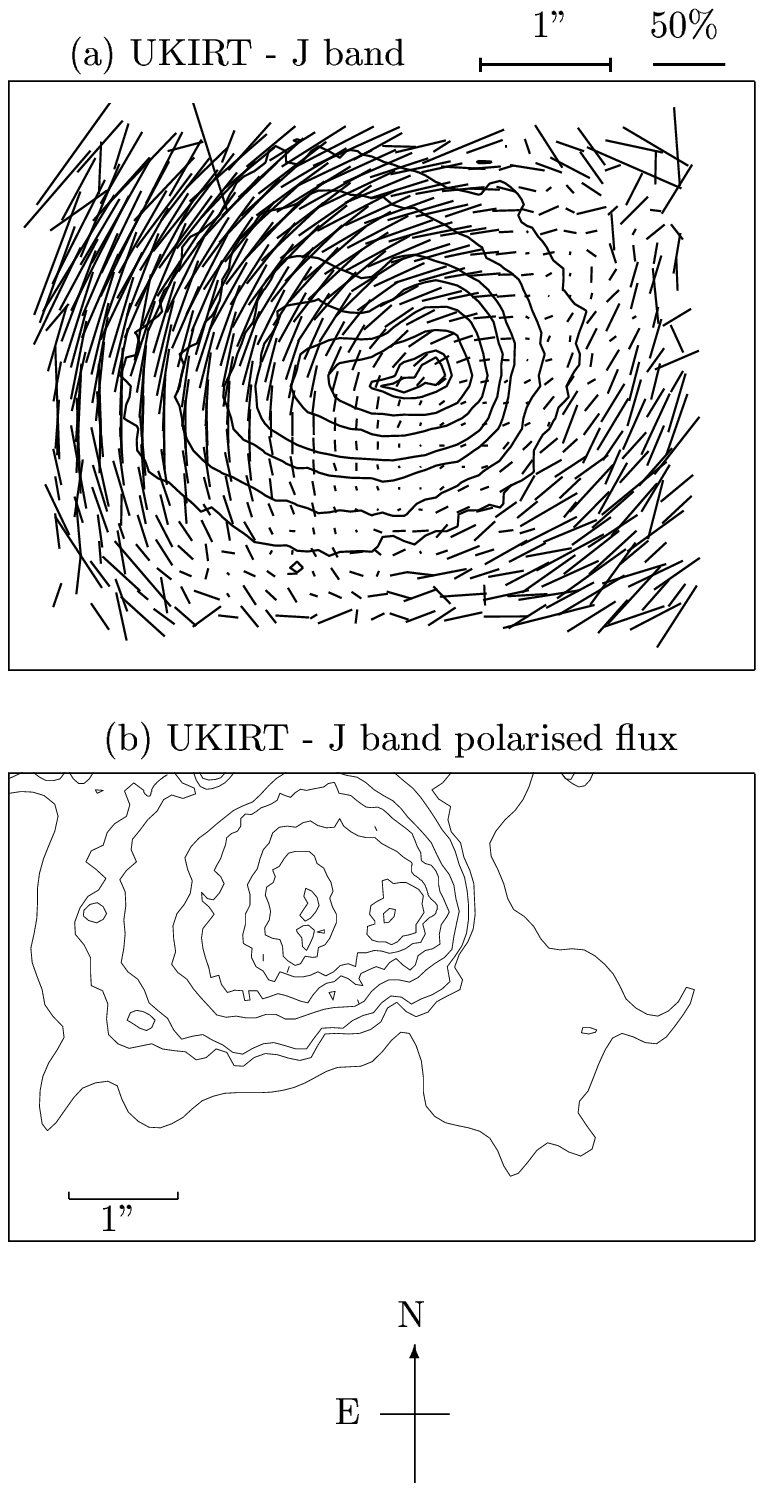}}

\end{picture} 
\end{center}
\vspace{-2cm}

Figure 3. H band data (left panels). (a) UKIRT image and polarisation map; (b) UKIRT polarised
flux image; (c) Subaru polarised flux image, covering a larger field of view. The UKIRT polarisation
vectors are 2$\times$2 binned as in Figure 2. Contour levels are normalised to unity at the peak. 
The levels in (a) are: 0.0125, 0.025, 0.05, 0.1, 0.2, 0.3, 0.5, 0.7, 0.8, 0.9, 0.98. Contour levels in 
(b) are 0.05(2), 0.075(2), 0.1, 0.2, 0.3, 0.5, 0.7, 0.9, where the '(2)' indicates contouring of 
2$\times$2 binned pixels. Levels in (c) 0.015, 0.025, 0.05, 0.075, 0.1, 0.2, 0.3, 0.5, 0.7, 0.9.

Figure 4. J band data (right panels). (a) UKIRT image and polarisation map; (b) UKIRT polarised
flux image. The UKIRT polarisation vectors are 2$\times$2 binned as in Figure 2. Contour levels 
are normalised to unity at the peak. The levels in (a) are: 0.05, 0.1, 0.2, 0.3, 0.5, 0.7, 0.9, 0.95.
Levels in (b) are 0.039(4), 0.07(2), 0.1(2), 0.2(2), 0.3, 0.5, 0.7, 0.9, where the '(n)' indicates 
contouring of n$\times$n binned pixels.

\end{figure*}
Weintraub et al.(1995) argued that
the central flux peak is not in fact the protostar and that it is dominated by scattered
light even in 0.4 arcsec resolution observations such as the speckle data of Beckwith et al.(1989).
This was motivated partly by the apparent offset of the K band flux peak from the focus of the 
centrosymmetric vectors. (The focus is not a unique position since the centrosymmetry is imperfect 
but Weintraub et al. were able to determine a best average focus location). However, this offset may
simply be due to the much lower spatial resolution of the Weintraub et al.(1995) measurement 
(using a camera with a 0.73 arcsec pixel scale), which might well have caused the flux peak to be 
dominated by scattered light. In addition, Monte Carlo modelling shows that the focus of the 
centrosymmetric vectors is unlikely to be coincident with the protostar in systems where the disc 
axis is inclined to the plane of the sky, (see LR98 and comment by Weintraub et al. 1995). 
In view of the later 0.2 arcsec resolution observations of Close et al.(1997) and the L band data of 
LR98 (which showed a very bright, essentially unpolarised point source at 0.2 arcsec resolution and
little nebulosity) we therefore argue that the protostar is observed directly in high resolution near 
infrared observations, with the precise wavelength depending on the spatial resolution.
Spectropolarimetry by Kobayshi et al.(1999) from 0.9 to 4.2~$\mu$m showed a steep
decline in polarisation with wavelength (using large apertures) which also suggests the
emergence of a point source. 

A component of emission from hot dust very close to the protostar may also contribute to the central
point source. Close et al. estimated from the weakness of NaI absorption at 2.21~$\mu$m that hot dust 
from the (spatially unresolved) innermost part of the accretion disc contributes 75\% of the 
point source flux at K band, veiling the infrared spectrum observed by Greene \& Lada (1996). However 
this does not take into account the low surface gravity of protostellar atmospheres, relative to field 
dwarfs of similar spectral type (type K). Sources with low gravity have weaker NaI absorption in the 
optical and the infrared (eg. Luhman et al.1998; Lucas et al. 2001) so the contribution from hot 
dust may be much less than 75\%, though it is hard to quantify.

\subsection{Polarised flux data}

The structure of the system cannot easily be inferred from the irregular nebulosity of the total flux
images (Figure 2(a,c), Figure 3(a) and Figure 4(a)). However, the polarised flux 
data (Figure 2(b,d), Figure 3(b,c), Figure 4(b)) suppresses the weakly polarised point 
source and reveals the receding lobe of the circumstellar envelope to the southwest and a 
dark lane of dense matter in between. The orientation of the plane of the disc and
envelope inferred from the dark lane is 136$^{\circ} \pm 8^{\circ}$ (east of north) and is 
broadly consistent with the value of $125^{\circ} \pm 10^{\circ}$ derived from millimetre data, 
see Section 5.1. The receding lobe is quite prominent and well defined in at K band, still more 
prominent at H band, but quite faint and diffuse at J band. This strong wavelength dependence, 
combined with the offset of the peak in total flux at J band, is a powerful constraint on the 
envelope's vertical density gradient and the extinction law, as shown in section 5. The receding lobe 
appears less prominent in the Subaru K band polarised flux data than the UKIRT data only because
the central point source is more highly polarised in the Subaru data (see section 3.2), so that the 
receding lobe and the dark lane appear at lower contrast. The polarisation of the receding lobe is 
the same in both datasets, and increases rapidly with distance from the protostar.

Other noteworthy features of the polarised flux data are: 

\begin{enumerate}
\item a broad peak (Source B) in the nebulosity located 1.3 arcsec east of the protostar, 
interpreted as asymmetrically distributed matter within the approaching lobe of the outflow cavity 
(see section 5)

\item a curved filament of nebulosity extending north from the vicinity of the protostar, seen in 
Figure 3(c). The filament is also seen in more recent non-polarimetic Subaru imaging data taken with 
AO (Tamura et al. 2004), and the outer part of it was seen in some 
previous imaging 
(Close et al. 1997; Takami et al. 1999). The filament is interpreted as the northern limb of the cavity wall 
associated with the approaching lobe of nebulosity. The eastern limb is not seen because it is
obscured by the bright dust in the cavity observed as Source B (see models in Section 5). 
\end{enumerate}

A non axisymmetric distribution of cavity dust is not particularly surprising since 
the orientation of the jet and outflow shows signs of varying over time (Close et al.; Mundt et al).
Moreover, Welch et al.(2003) report the detection of a companion millimetre continuum source 
with significant mass (M $>10^{-3}~$M$_{\odot}$) at P.A.$\approx 315^{\circ}$ some 0.6 arcsec 
(84~AU) from the centre of main accretion disc (radius 0.45 arcsec or 63~AU). The nature of the 
companion source is unclear (a dust cloud or a binary companion are both possible) but it may well 
influence the structure of the outflow cavity. The millimetre companion is not seen in the IR.

\subsection{Polarisation maps and magnetic field structure}

The most interesting feature of the polarisation maps is that the polarisation vectors associated 
with the protostar are inclined by between 30$^{\circ}$ and 45$^{\circ}$ to the 
disc plane. The 0.16-0.17 arcsec
pixel scales used in Figures 2, 3 and 4 oversample the spatial resolution so several polarisation 
vectors around the protostar with very similar orientation are seen in the Subaru and UKIRT K band 
polarisation maps and the UKIRT H band map (the Subaru H band vector map suffered from imperfect 
tracking, which obscured small scale structures and is therefore not shown). Figure 5(a-c) shows
a more detailed view of the central regions, with the UKIRT data unbinned (0.081 arcsec pixels)
and the Subaru data $4 \times 4$ binned (0.088 arcsec pixels).

\vspace{-4cm}
\psfig{file=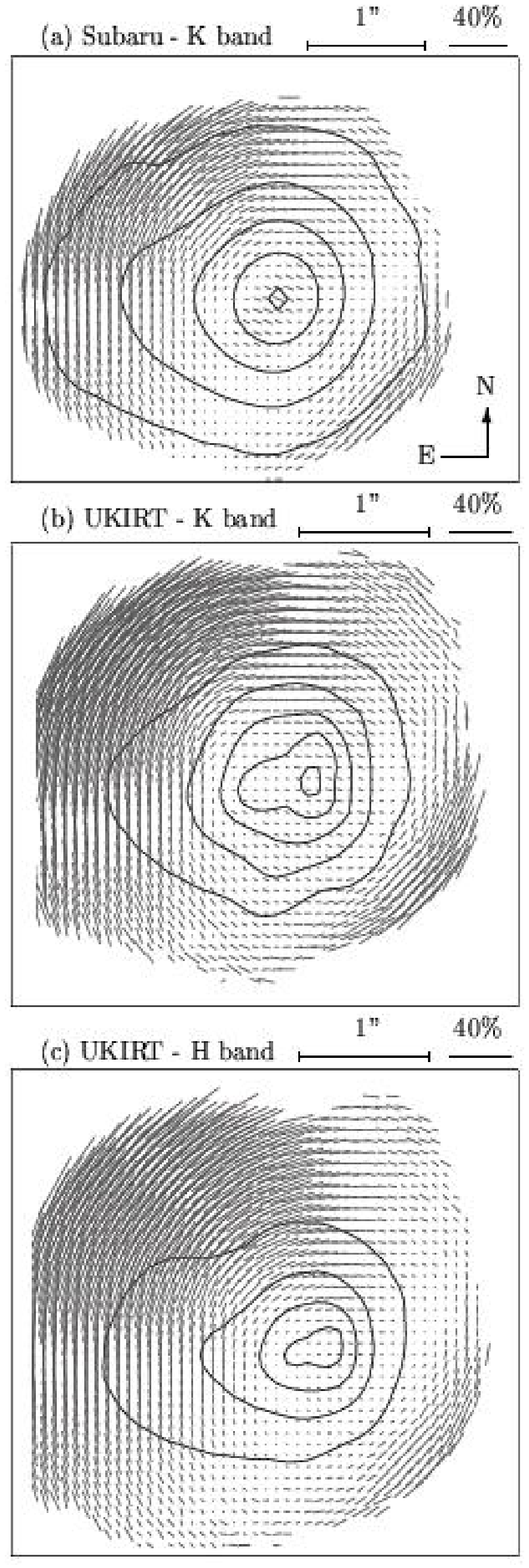,clip=,width=8.3cm}

\vspace{-4mm}
Figure 5(a-c). The central region in close up. (a) Subaru K band data ($4 \times 4$ binned)
to a 0.088 arcsec pixel scale. (b) UKIRT K band data. (c) UKIRT H band data. UKIRT data
are unbinned, giving a 0.081 arcsec pixel scale. The vectors asscoiated with the central
source appear somewhat inclined to the disc axis. The orientation of the disc axis 
can be inferred from bipolar structure of the vector pattern in each plot, although the
true disc orientation measured at millimetre wavelengths is a little different (see text).

\begin{figure*}
\begin{center}
\begin{picture}(200,300)

\put(0,0){\includegraphics{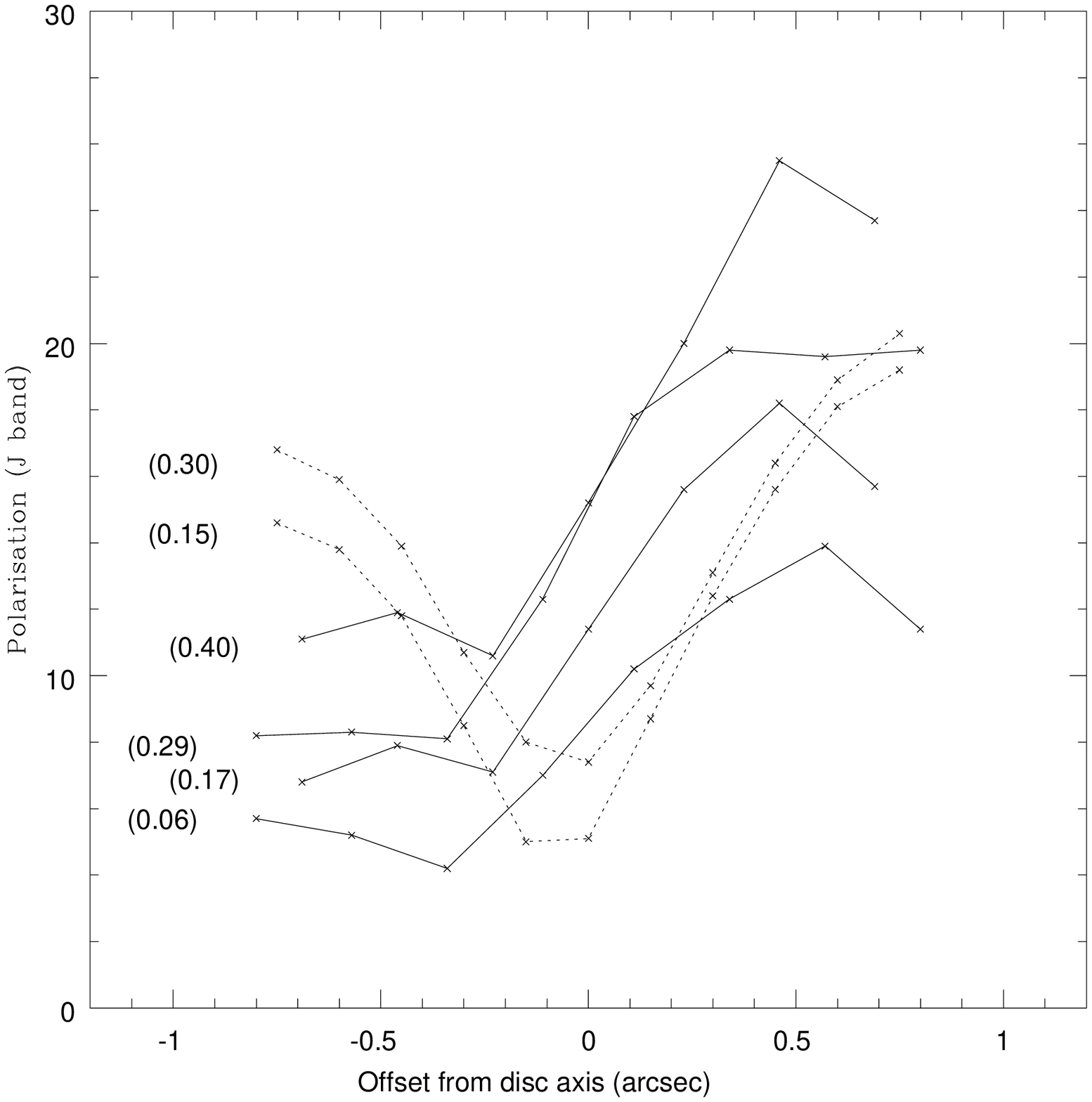}}

\end{picture} 
\end{center}
Figure 6. Asymmetry of J band polarisation. The plot shows several cuts across the disc axis
in the approaching lobe: solid lines are data, dotted lines are model results. The location of each
cut in arcseconds north east of the J band flux peak is marked. Polarisation in both the models
and the data is higher to the right of the axis (positive offsets) and increases with distance 
along the axis. The polarisation gradient across the axis in the models is due to 
dichroic extinction of scattered light (see text). The effect in the data is greater and more 
spatially extended than in the models, suggesting that additional factors may be at work, eg.
the structural asymmetry of the nebula. A P.A. of 45$^{\circ}$ is adopted for the disc axis in this
plot to avoid rotating and resampling the data but this P.A. has some uncertainty (see text).
Cuts using a P.A. of 35$^{\circ}$ (not shown) agree better with the rise in polarisation at 
negative offsets seen in the models but show a declining polarisation at $X>1$~arcsec.

\end{figure*}
The small group of inclined vectors associated with the central source is easily fitted using the 
unresolved protostar in the models in section 5. The position angle and 
degree of polarisation observed by UKIRT and Subaru around the protostar are a little different 
(see Table 1) but the structure is qualitatively the same in both datasets and the difference is 
probably due to measurement error. The uncertainties in the polarisation 
measurements for the small aperture used in Table 1 are likely to be dominated by changes in the image
profile between waveplate positions rather than photon shot noise or registration error. The models 
described in section 5 show that the measured polarisation at K band in a 0.4 arcsec aperture
is insensitive, to first order, to the slight difference in spatial resolution of the two datasets.
The uncertainties in Table 1 are estimated from internal variations of the measurements in subsets of the 
data: UKIRT had 3 independent cycles of the waveplate position and Subaru had 4 independent cycles.
The individual UKIRT K band cycles had values of 3.39\%, 3.35\% and 3.86\% for the central source.
These average to a slightly lower value due to slightly different position angles in each subset.   
The UKIRT data has higher precision than Subaru due to the dual beam system of IRCAM. In theory the
stable focus offset should not affect the measurement. However it is possible that 
the Subaru measurement is in fact more accurate because of the UKIRT focussing problem so both 
datasets were taken into account when deriving uncertainties in model parameters.

The typical polarisation in both lobes of the nebula is very similar in the UKIRT and Subaru datasets.\\

If we adopt P.A.=35$^{\circ}$ as the orientation of the disc axis (this being the orientation of the 
minor axis of the millimetre disc observed by Mundy et al.(1996); Wilner et al.(1996)) then the
data in Table 1 indicate that the polarisation vectors associated with the point source are inclined 
by between 32$^{\circ}$ and 43$^{\circ}$ to the disc axis. The polarised flux maps suggest 
P.A.=46$^{\circ} \pm 8^{\circ}$ for the disk axis (see 
above) and the orientation of the optical jet (Mundt et al. 1990) suggests P.A.$\approx$ 50$^{\circ}$. 
These values (and casual inspection of Figure 5(a-c)) would suggest a slightly smaller inclination to the
disc axis. However they are derived from or influenced by the distribution of matter in the outflow 
cavities, which may not be ejected along the disc axis.\\

\begin{center}
\textbf{Table 1 - Polarisation of the central point source}
\begin{tabular}{lcc}
 & & \\
Data & P.A. & Polarisation\\ \hline
UKIRT K band & 79.7$^{\circ}$ & $3.3\% \pm 0.3\%$\\
UKIRT H band & 77.4$^{\circ}$ & $3.7\% \pm 0.3\%$\\
Subaru K band & 68.5$^{\circ}$ & $6.8\% \pm 1.7\%$\\
\end{tabular}
\end{center}
\small
Note: polarisation is measured in a 0.4 arcsec diameter aperture centred on the point source.\\

\normalsize

The polarisation of the point source is interpreted as the result of dichroic extinction of light 
from the protostar by aligned non-spherical grains. This interpretation is supported by the 
observation of a distortion of the centrosymmetric pattern between at distances from 0.5 and to 1.5
arcsec from the point source which is consistent
with a dichroic extinction pattern that is superimposed upon the centrosymmetric pattern produced by 
scattering. The apparent dichroic extinction seen at H and K produces polarisation at 
P.A.$\approx 79^{\circ}$ (see Table 1), which would be expected to add to the scattering-induced 
polarisation at locations north of the protostar, but subtract from the 
scattering-induced polarisation to the east of the protostar. Cuts across the disc axis in the 
approaching lobe are shown in Figure 6, illustrating that there is a polarisation gradient across the 
axis at J band.

This behaviour is observed at J band on scales up to 1.5 arcsec ($\sim 200$~AU) from the flux peak 
(which is offset slightly from the protostar, see section 5), and is attributed to dichroic 
extinction of scattered light passing through the dense inner parts of the envelope. It is possible 
that the polarisation gradient across the disc axis relates at least in part to structural 
asymmetries of the nebula however. The same distortion
is apparently observed at H and K band in the data from both telescopes but that might be 
simply due to contamination by direct flux from the bright nearby protostar (the central point source)
in the wings of the image profile.
Light which is scattered within $\sim 100$ AU of the protostar will experience a dichroic extinction 
similar to that of direct flux from the photosphere or hot accretion disc. This interpretation and the 
information derived about the magnetic field by Monte Carlo modelling are discussed further in 
Section 5. 

The observed polarisation structure is unusual, though only a few YSO have been subject to
polarisation measurements at this high spatial resolution.
Class I YSOs commonly exhibit a region of aligned vectors near the projected location of the 
protostar. These are usually aligned with the disc plane, and are often interpeted in terms of 
polarisation dominated by double scattering, due to high optical depth on the line of sight to the 
central source (Whitney \& Hartmann 1993; Bastien \& Menard 1988). Aligned vectors
have occasionally been observed in less embedded Class I sources similar to HL Tau, where the 
protostar is seen directly through a lower optical depth ($\tau < 6$). For example, LR98 
observed a 'polarisation disc' aligned parallel to the disc plane of TMR-1, which 
has a bright central flux peak. 

On larger scales the polarisation pattern of HL Tau is approximately centrosymmetric, though the
centrosymmetry is imperfect in places due to the effects of both double scattering and 
the finite width of the image profile: polarisation measured in areas of faint nebulosity is
contaminated by the polarisation signal from adjacent areas of brighter nebulosity. 
The maximum polarisation measured in the outer nebulosity in a 0.24 arcsec aperture is 80\% in the 
K and H bands in the UKIRT data, and 75\% in the Subaru K band data. The maximum polarisation 
measured in the UKIRT J band data is 75\%.



\section{Monte Carlo light scattering codes}
 
Three Monte Carlo codes were used to investigate the 3-D structure of the envelope
around HL Tau and the structure of the magnetic field, which is inferred from the 
direction of grain alignment. The first code, 'range\_asph.f' is a 2-D code used to
model YSOs which are symmetric about the axis of the accretion disc and the envelope
(axisymmetric). The code is described in Lucas (2003). It incorporates scattering by 
oblate spheroids which are perfectly aligned with an arbitrary magnetic field structure. 
This is an extension of the code used by LR98 ('range.f') to model scattering by spheres, with the
non-spherical particles allowing us to include the effects of dichroic scattering and 
dichroic extinction on model images and model polarisation maps. This code has been
qualitatively bench tested against that of Whitney \& Wolff (2002). The T-matrix
code ampld.f of Mishchenko, Hovenier \& Travis (2000) is used to generate the large look up
tables required to calculate both the polarisation after each scattering of a photon and the
phase function from which the scattering direction is sampled by rejection sampling.
The range\_asph.f code was used with a uniform axial magnetic field to constrain the basic 
parameters of HL Tau, such as system inclination, optical depth and the approximate axis ratio 
of the oblate grains, before moving to a 3-D analysis.

The second code, 'whale.f', is a very simple extension of range\_asph.f to 3-D, which was
used to model the non-axisymmetric matter in the cavity of HL Tau. The great 
drawback of such a 3-D code in which the photons are randomly scattered is that only a 
small fraction of the photons used by the code are output into the direction of observation. 
In 2-D axisymmetric codes all photons can be assumed to exit at the correct azimuthal angle 
for observation, and output photons are binned only by polar angle (corresponding to 
different system inclinations). Run times of 7 hours were required with 8 CPUs on a fairly 
fast parallel SGI Origin 2000 computer to produce rather poor quality model images from an 
initial 20 million 
photons originating at the photosphere. Despite this drawback whale.f is useful to explore 
the range of possible orientations of the system, since the output can be used to make images 
for any direction of observation. In range\_asph.f and whale.f the photons at first have no chance 
for absorption during a scattering event. Instead the more efficient process of multiplying the
Stokes vector of each photon by a weighting factor is used. The weight is given by:\\

(1) W = $\omega^N$\\

where $\omega$ is the albedo and N is the number of times the photon has been scattered. 
Absorption of photons is not permitted until W$<10^{-3}$, whereupon the weight is held constant and 
absorption or scattering occurs in the natural random manner.

Following discussions with K. Wood (private comm.) a third code was written, 'shadow.f',
which uses the technique of forced scattering to permit much faster 3-D modelling. The forced
scattering technique is described by Cashwell \& Everett (1959); Yusef-Zadeh, Morris \& White (1984) 
and Wood \& Reynolds (1999), for application to scattering by spheres. Photons propogate along 
random multiple scattering paths until they exit the system, as in whale.f. However, every time a 
photon scatters, 
a 'shadow photon' is generated at the scattering location and propogates into the user specified 
observation direction. The shadow photons penetrate any optical depth and never scatter, so they are 
weighted according to the probability of such an event. Each shadow photon therefore has a weight:\\

(2) W = $\omega^N$ P(D,$\gamma$) e$^{-\tau}$\\

where N is the number of times the parent photon has been scattered; e$^{-\tau}$ is the 
probability of a photon travelling the corresponding optical depth $\tau$ to the edge of 
the system without scattering or absorption; and P(D,$\gamma$) is the probability of 
scattering through the corresponding deflection angle D and azimuthal angle $\gamma$. The azimuthal
deflection angle
$\gamma$ is defined as a left handed rotation about the Poynting vector of the parent photon, 
with zero at the incident polarisation position angle. 
P(D,$\gamma$) is a function of the grain mixture (size distribution and shape), grain 
orientation and the (IQUV) polarisation state of the incident photon. Shadow.f is far more 
efficient than whale.f, with speed closer to a 1-D code than a 3-D code. Calculation times of only 
6 minutes with 8 CPUs were required to generate high quality models, using only 
10$^5$ photons (with one caveat, see below). Forced scattering codes
such as this have different statistical properties than fully stochastic codes. They provide
a more 'democratic' spread of model image quality due to the penetration of high optical
depth by shadow photons. The signal to noise in the obscured receding lobe of a bipolar 
nebula is similar to that of the bright approaching lobe. This code was used to refine 
models of the following: (i) the magnetic field structure in the envelope; (ii) the matter 
distribution in the envelope and the bipolar cavity; and (iii) the near infrared extinction law 
and the grain axis ratio.

Shadow.f was coded by adding a single subroutine to whale.f to handle the shadow photons. 
Two complications are worthy of mention. The first is that the output flux of shadow photons 
is not normalised relative to the direct (unscattered) flux from the protostar. This
is significant in cases like HL Tau where the protostar is directly observed through a 
modest optical depth. The normalisation was handled by causing shadow.f to produce the 
traditional stochastic output of whale.f as well as the forced output. The ratio of direct 
flux to total scattered flux in the stochastic output was then used to determine the 
appropriate direct flux in the shadow photon images. A reasonably precise normalisation 
(better than 10\% precision) required longer runs with 10$^6$ photons.

The second complication of shadow.f is a difference between the phase function for forced 
scattering and that for stochastic scattering. The phase function for forced scattering is the 
probability P(D,$\gamma$) per unit solid angle, whereas the phase function for stochastic 
scattering is P(D,$\gamma$)$sin(D)dDd\gamma$, incorporating the amount of solid angle in a bin
of width $dDd\gamma$ at deflection angle $D$. The need to remove the $sin(D)$ term was discovered
during exhaustive bench testing to provide precise quantitative agreement between outputs from 
shadow.f and whale.f. The testing used a homogeneous sphere with a uniform magnetic field, a 
homogeneous sphere with a pinched and twisted field, and a model of HL Tau. 

The phase function used in shadow.f is:\\\vspace{2mm}
(3)$P(D,\gamma) = \{0.5(1-LP) 
(|S_{11}|^2 + |S_{12}|^2 + |S_{21}|^2 + |S_{22}|^2)$\\\vspace{2mm}
$+ LP[(|S_{22}|^2 + |S_{12}|^2)sin^2(\gamma) + (|S_{11}|^2 + 
|S_{21}|^2)cos^2(\gamma)$\\\vspace{2mm}
$\hspace{1.2cm} + sin(2\gamma)Re(S_{11}S_{12}^{*} + S_{22}S_{21}^{*})]\\\vspace{2mm}
-CP(Im(S_{11}S_{12}^{*} - S_{22}S_{21}^{*}))\}
/T(\alpha_0,\beta,LP,CP)$

where the S$_{ij}$ terms are S$_{ij}(\alpha,\beta,D,\gamma)$, the (complex) elements of the 
2$\times$2 amplitude matrix used to calculate the scattered components of the electric field 
parallel and perpendicular to the scattering plane (eg. Mishchenko et al.2000). As described 
in Lucas (2003), $\alpha$ and $\beta$ are the azimuthal and polar grain orientation angles in 
the frame of the incident photon. LP and CP are the fractional linear polarisation and 
circular polarisation of the incident (parent) photon. To reduce the size of the look up 
tables for the S$_{ij}$ arrays, the photon azimuthal deflection angle $\gamma$ is set to 
zero in the S$_{ij}$ elements and $\alpha$ (the grain azimuthal orientation angle) is 
calculated relative to the azimuthal deflection. The numerator in Eq.(3) is the scattered flux per 
unit solid angle. The denominator $T(\alpha_0,\beta,LP,CP)$ is used to normalise the 
scattered light intensity as a probability function. $T(\alpha_0,\beta,LP,CP)$ is the sum of the 
numerator in Eq.(3) evaluated over all possible scattering angles ($D,\gamma$), for the combination of 
($\alpha_0,\beta,LP,CP$) appropriate to the parent photon. The term $\alpha_0=\alpha-\gamma$. It 
represents the azimuthal orientation of the grain perceived by the incident photon, relative to the 
polarisation plane, correcting for the fact that $\alpha$ is defined relative to $\gamma$. (Note 
that $\alpha$ rather than $\alpha_0$ was erroneously used in Lucas (2003) in the normalisation term 
'$IMAX(\alpha_0,\beta,LP,CP)$' for the phase function in the stochastic code range\_asph.f. However 
testing indicates the error was too subtle to have any measurable effect on model results).\\

{\bf Image Convolution}\\

The (IQUV) maps produced by the above codes have to be convolved with a point spread function
(PSF) which is the best available match to the image profile yielded by the telescope. The PSF for 
Subaru/CIAO
was determined by fitting a field star observed shortly after HL Tau with a moffat profile in the IRAF
software package. No suitable reference data was obtained during the UKIRT Service observations
of HL Tau on the same night, so the rebinned image profile based on similarly out of focus data from 
the previous night (see right panel of Figure 1) was used as the basis for the convolution kernel. The 
effective
spatial resolution on that night was approximately 0.65 arcsec. However, after convolving with the 
J, H and K band models this resolution was slightly too poor to reproduce much of the fine structure 
seen in the data, such as the double peaked profile of the J band polarised flux data in Figure 4(b). The 
convolution kernel was demagnified by varying amounts to give effective resolutions ranging from 0.4 to 
0.65 arcsec to assess the true resolution. Models with resolutions ranging from 0.4 to 0.6 arcsec were 
found to fit the data, although 0.5 to 0.6 arcsec resolution is most likely due to the defocussing.
A kernel with a resolution of 0.55 arcsec was therefore adopted and used in the models
of the UKIRT data. The uncertainty in the true spatial resolution of the UKIRT J band data
increases the uncertainty in the fitted J band optical depth and the derived extinction law in 
section 5.

\section{Modelling HL Tau}
\subsection{Initial assumptions and prior data}

The protostar is assumed to be surrounded by a physically thin but flared accretion disc in 
hydrostatic equilibrium (Shakura \& Sunyaev 1973), described by:\\

\hspace{-4mm}$\rho_{disc}=\rho_{0}(r/R_{*})^{-15/8}exp[-0.5(z/h)^{2}]; \hspace{2mm} R_D>r>R_{min}$

(4) \hspace{0.9cm} $= 0; \hspace{1mm} r< R_{min}$

\hspace{1.4cm} $= 0; \hspace{1mm} r> R_D$\\ 

where $r$ and $z$ are cylindrical coordinates defined by the disc plane and $h$ is the disc scale 
height, given by:\\

\hspace{-4mm}(5) $h = h_{0}(r/R_*)^{9/8}$;  \hspace{2mm} $h_0 = R_{*}/75$\\

where $R_{*}$ is the stellar radius, set at 2~$R_{\odot}$ (or $\sim 0.01$~AU).
The small dust grains observed in the near infrared are assumed to be well mixed with the gas. The 
accretion disc may contribute to the K band emission, providing a different angular distribution of 
photons than the spherical protostar. The dust disc is assumed to have an inner hole within radius
$R_{min}$ due to disc disruption or dust sublimation. Default parameters of $R_{min} = 2R_*$,
$R_D=150~AU$ and zero thermal contribution from the disc were adopted. In practice the structure of the
accretion disc and any thermal flux have very little effect on the model output, which is dominated by 
the larger scale envelope and the bipolar cavity.

For the larger scale envelope two density structures were investigated, described by Eqs.(6) and (7):
\\

\hspace{-6mm}
$\rho_{env} = C/R^{1.5} [1/((|z|/R)^{v} + 0.05)]; \hspace{4mm} R_{cav}<r<1300~AU$\\
(6) \hspace{3mm}        $  = 0; \hspace{2mm} r<R_{cav}$\\ 

where $R=\sqrt{r^2+z^2}$, $C$ is a free parameter governing the optical depth and the power law index $v$ is a parameter 
describing the vertical gradient of the envelope. $v=0$ correesponds to spherical symmetry, while
increasing to positive values causes progressively more flattening of the envelope. $R_{cav}$ is the
radius of the evacuated conical cavity in the disc plane, for which an opening angle 
$\theta_c=45^\circ$ to the disc axis was adopted in the initial model.
Eq.(6) is a simple, empirically derived formalism, adopted following the discovery that the $R^{-1.5}$ 
power law seems to extend throughout the envelope in at least some YSOs (LR98).\\ 

\hspace{-4mm}
$\rho_{env} = \frac{\dot{M}}{4\pi} \frac{1}{(GM)^{1/2}} \frac{1}{R^{3/2}} \frac{1}
{(1 + \mu/\mu_{0})^{1/2}} \frac{1}{(\mu/\mu_{0} + 2\mu_{0}^{2}c_{r}/R)}$;\\
$\hspace{4mm} R_{cav}<r<1300~AU$\\
(7)\hspace{3mm} $  = 0; \hspace{2mm} r<R_{cav}$\\ 

where M is the mass of the protostar, \.{M} is the accretion rate, $\mu = |z|/R$  and 
$\mathrm{\mu_{0}=f(r/c_{r},\mu)}$ (see Whitney \& Hartmann 1993). Eq.(7) is the inner solution
of Terebey, Shu \& Cassen (1984) for the inside out collpase of a rotating singular isothermal sphere,
also derived by Ulrich (1976). Eq.(7) differs from Eq.(6) primarily in the radial density profile:
at radii much greater than the 'centrifugal radius' c$_{r}$, the density profile varies as R$^{-3/2}$, 
but at smaller radii the density varies more slowly, as R$^{-1/2}$ for $\mathrm{R << c_{r}}$. The 
density distribution is nearly spherical at $\mathrm{R >> c_{r}}$ but becomes flattened at 
smaller radii. The outer limit to the system is set at 1300 AU in both formalisms, since this is the 
approximate extent of the observed millimetre and infrared structures (eg. Hayashi et al.1993).

In the initial model the cavity structure was assumed to be conical and perfectly evacuated, with a 
radius of 25~AU in the disc plane, with an opening angle of 90$^{\circ}$.
 
Two important parameters in the model fit are the system inclination and the optical depth to the 
central star. The most direct mean of determining the inclination is from the elongation of the disc 
seen in the millimetre continuum. Those observations which have clearly resolved the disc with a P.A.
near 125$^{\circ}$ imply inclinations $i = 60^{\circ +14}_{-16}$ (Mundy et al.1996 at 2.7~mm), 
$i =62^{\circ +9}_{-10}$ (Wilner et al. 1996 at 7~mm).
High resolution $^{13}$CO maps by Hayashi et al.(1993) imply $i \approx
67^{\circ}$, though the inference from CO data is less direct. Close
et al.(1997) adopted $i=67^{\circ}$ in their Monte Carlo model. By
contrast, Menshchikov et al.(1999, hereafter MHF99) prefer a much
lower value ($i=43^{\circ}$), which  follows from their adoption of a
model with a physically thick inner torus on a 100 AU scale, as
opposed to our more conventional physically thin inner accretion disc
on a similar scale. The thick torus model put forward by MHF99 for
Class I YSOs is thought provoking, even though simplified by its essentially 1-D nature 
(see Menshchikov \& Henning 1997 for the computational 
details). MHF99 modelled the entire optical to centimetre waveband SED of the system with a 
4-component, spatially segregated dust model. They also modelled low resolution imaging 
polarimetry with the Monte Carlo method, though with limited quantitative success. 
This model of a thick inner disc which is not yet in
hydrostatic equilibrium cannot be entirely ruled out but we consider
such a thick torus (with the surface rising linearly at 45$^{\circ}$
from the disc plane) to be unlikely for the following reasons: (i) the high elongation of the 
disc observed in the millimetre waveband on several occasions argues strongly for a large
inclination of the  system, almost certainly with $i>45^{\circ}$;
(ii) their fit requires that the line of sight passes very close to
the cavity wall as viewed in the near IR, which is something of a
coincidence; (iii) the observation of an unresolved central source at 2~$\mu$m here 
and in Close et al.(1997) suggests that the protostar is observed directly, 
contrary to their model; and (iv) millimetre wave mapping and Monte Carlo modelling of 
IRAS 04302+2247, another Class I YSO in Taurus believed to be at a similar
evolutionary stage, observed at i=90$^{\circ}$ (eg. Wolf et al. 2003;
LR97) has preferred flatter disc models. LR97 discounted thick discs
on a 100 AU scale because these were detectable and very prominent in
high resolution Monte Carlo maps but not seen in the data.

The optical depth to the central source is not known a priori. Close et al.(1997) 
provide point source fluxes at J, H and K, though with significant uncertainty at J band
due to the difficulty of precisely subtracting the scattered light contribution. 
If the extinction law is known the K band optical depth $\tau_K$ can be derived from 
the (J-H) colour excess, assuming this is almost entirely caused by extinction. 
We take the (J-H) colour of the photosphere to be 0.58, using the derived temperature 
T$_{eff} = 4600$~K of Close et al. and the colours of Tokunaga (2000). (The colours 
are the same for Class III and Class V stars, and are only a weak function of 
T$_{eff}$). Hence E(J-H)=2.59. Adopting the $\lambda^{-1.61}$ near infrared extinction law 
of Cardelli, Clayton \& Mathis (1989) leads to $A_K=2.89$~mag or $\tau_K=2.66$. Adopting the shallow 
extinction law suggested by LR98 for Class I YSO envelopes in Taurus leads to $A_K=5.63$~mag or 
$\tau_K=5.18$. Values between these two extremes were explored in the simulations.

The magnetic field was initially assumed to be uniform and parallel to the disc axis. The more 
complicated structures required to fit the polarisation maps in the vicinity of the protostar are 
described in section 5.2.2.


\subsection{The successful model}

The parameters of the successful model are summarised in Table 2 and it is illustrated in 
Figure 7 with the disc axis oriented vertically. The features which the model fits are:

\begin{figure*}
\begin{center}
\begin{picture}(200,680)

\put(0,0){\includegraphics{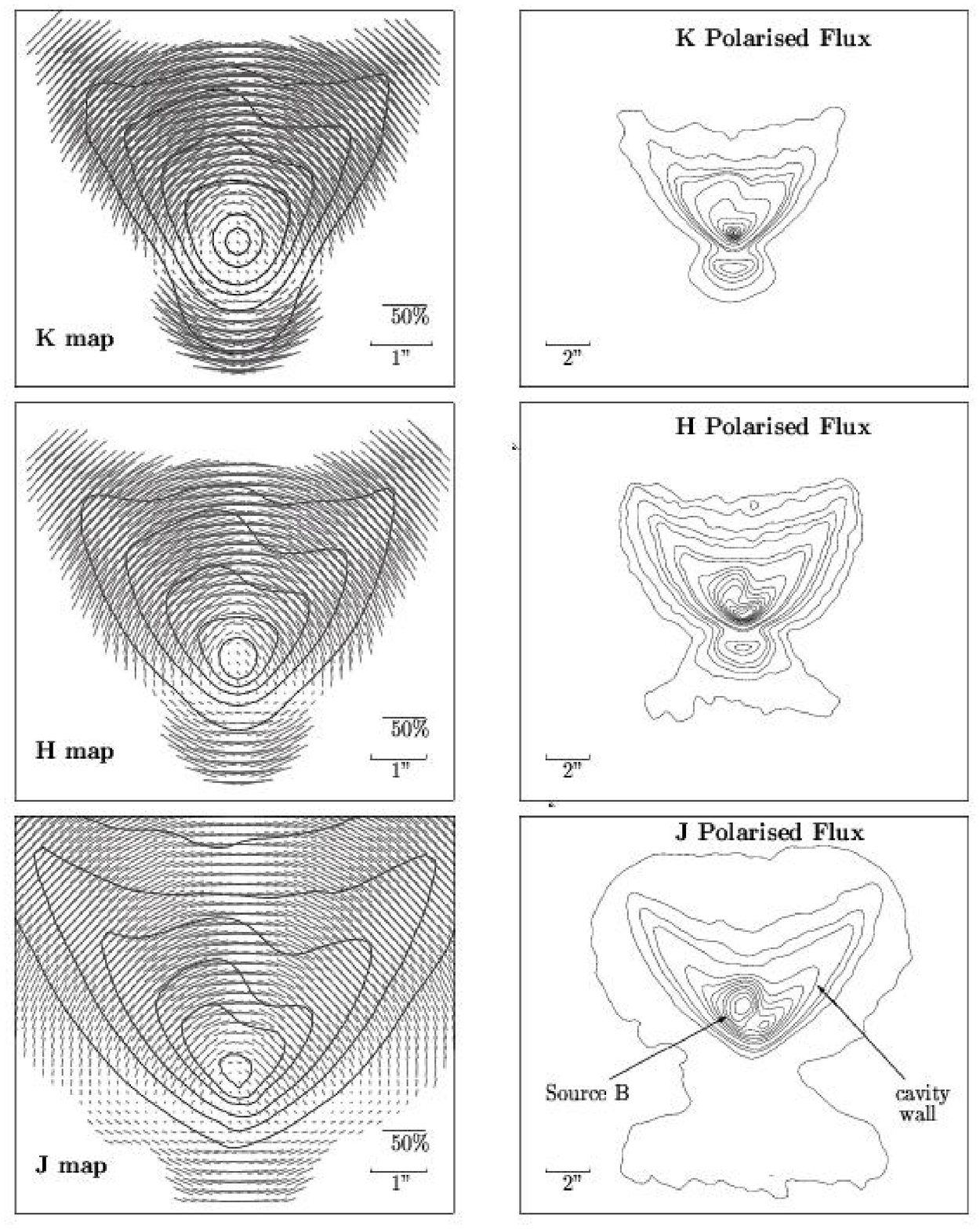}}

\end{picture} 
\end{center}

\small
\vspace{-3.2cm}
Figure 7. Successful model. (top) K band; (middle) H band; (bottom) J band. Left panels show
the model images and polarisation maps, including only the central regions for clarity.
The pixel scale is 0.15 arcsec, corresponding to 21 AU in the system. 
Right panels show polarised flux images of the whole system with similar dynamic range to the data. The 
K and H band models are smoothed to the 0.4 arcsec resolution of the Subaru data, while the J band 
model is smoothed to the 0.55 arcsec resolution of the UKIRT data (see section 4). 
All contours are normalised to unity at the peak. Contour levels in the polarisation maps areas follows. 
K band: 0.5, 0.16, 0.03, 
0.015, 0.0075, 0.0038. H band: 0.5, 0.25, 0.125, 0.0625, 0.031. J band: 0.9, 0.85, 0.43, 0.21, 0.11, 
0.053. The polarised flux images have the following contour levels. K band: 0.01, 0.02, 0.03, 0.04, 
0.05, 0.1, 0.2, 0.3, 0.4, 0.5, 0.6, 0.7, 0.8, 0.9. H band: 0.015(2), 0.02, 0.03, 0.04, 0.05, 0.08,
0.1, 0.2, 0.3, 0.4, 0.5, 0.7, 0.8, 0.9. J band: 0.01(2), 0.04, 0.07,
0.1, 0.2, 0.3, 0.4, 0.5, 0.62, 0.7, 0.8, 0.9. The '(2)' indicates 
contouring of 2$\times$2 binned pixels.   
\end{figure*}

\normalsize
\begin{enumerate}
\renewcommand{\theenumi}{(\arabic{enumi})}
\item Quantitatively, the ratio of polarised flux in the brightest part of the receding lobe to that 
at the central polarised flux peak, evaluated in the J, H and K bands. 
A 0.4 arcsec aperture was used for the measurements. The measured ratios in the UKIRT data are 0.065 
at K band, 0.11 at H band, and $\sim 0.05$ at J band. The ratios in the Subaru H and H band data
are the same after accounting for the higher polarisation of central flux peak in the Subaru data.  
The prominence, location and well defined boundaries of the receding
lobe in polarised flux could only be fitted with a component of dust in the receding lobe of the 
outflow cavity.

\item Quantitatively, the offset along the disk axis of the J band peak in total flux from the centre 
of the polarisation pattern, as defined by the polarisation nulls southwest of the flux peak.
The nulls are used since there is no clear 'polarisation disc' and no unique focus for the 
centrosymmetric vectors in the outer nebulosity.

\item Quantitatively, the radial profile of the K band flux image, along a cut extending along the 
disc axis from the protostar to the edge of the approaching lobe. This feature is fitted using the
Subaru data, for which the image profile was well defined by observation of a stellar standard
(see Section 4).

\item Quantitatively, the maximum degree of polarisation in the HL Tau nebula at J, H and K, evaluated 
in a 0.24 arcsec aperture.

\item Quantitatively, the position angle and percentage of the K band polarisation of the central
point source. The magnitude and structure of the polarisation vectors
around the protostar are also reproduced well in the H and J bands.

\item Qualitatively, the distortion of the centrosymmetric pattern within 100 AU (0.7 arcsec) of  
the protostar by dichroic extinction of scattered light, observed at J, H and K (see Figure 6). A
quantitative fit was not attempted owing to the uncertainties caused by sensitivity to the convolution 
function used to smooth the models.

\item Qualitatively, the presence of the broad scattering peak called Source B, seen in polarised flux 
at J, H and K and modelled with dust in the approaching lobe of the bipolar cavity. Source B is most
clearly seen in the J band model.

\item Qualitatively, the presence of the filament of nebulosity extending north of HL Tau in 
Figure 3(c), interpreted as a limb brightened cavity wall of the approaching lobe.

\item Qualitatively, the shape of the bipolar nebula in the polarised flux images. This is reproduced 
better by curved cavity walls (in the parabolic sense) than by a conical cavity profile. In addition
the modest indentation of the contours in the dark lane southwest of the protostar is better fitted by 
dust grains with a fairly high albedo ($\gtsimeq 0.3$ at H and K), since significant multiple 
scattering permits stronger illumination of the equatorial plane of the envelope, which is optically 
thick at low latitudes (see Figure 14 of LR98). 
\end{enumerate}

\vspace{1cm}
\begin{center}
\textbf{Table 2 - Parameters of the Successful Model}\\
\vspace{4mm}
\begin{tabular}{lcc}
Parameter & Value \\ \hline
System Radius & 1300 AU \\
Optical Depth to the protostar at J, $\tau_J$ & $7.10^{+1.31}_{-0.45}$\\
Optical Depth to the protostar at H, $\tau_H$ & $5.25 \pm 0.35$\\
Optical Depth to the protostar at K, $\tau_K$ & $3.75 \pm 0.50$\\
envelope vertical density gradient, v & $0.65 \pm 0.15$ \\
grain axis ratio, $gr$ & $1.015^{+0.015}_{-0.005}$ & \\
system inclination, i, & $66^{\circ}<i<71^{\circ}$ \\
cavity radius in disc plane, R$_{cav}$ & 25 AU \\
$^*$cavity opening angle, $\theta_c$ & 45$^{\circ}$ \\
$^*$cavity curvature parameter, $cc$ & $0.1^{+0.2}_{-0.1}$ \\ 
\end{tabular}
\end{center}
{\small * The cavity profile is described in Eq.(8). The opening angle is defined relative to 
the disc axis.

\normalsize
\subsubsection{Structure of the circumstellar matter}

Two modifications to the initial structural assumptions of Section 5.1 were required to produce this 
fit. First, dust in both lobes of the cavity (not merely within the cavity walls) was required to fit the 
prominent and well defined 
receding lobe seen in polarised flux (Figure 2(b,d); Figure 3(b,c)), and the diffuse Source B in the 
approaching lobe. For simplicity this dust is assumed to have the same properites as the dust in the 
envelope and the disc (see section 5.2.3).
The dust distribution used in the receding lobe is axisymmetric and located at $200<R<400$~AU. The
optical depth along the disc axis in the receding lobe is $\tau \sim 2$ at K band, which maximises
the lobe brightness. An $R^{-2}$, approximately 'constant flow' density distribution provides the best
qualitative match to the data. The dust in the approaching lobe has to be non-axisymmetric to reproduce
Source B and explain why
only one limb of the cavity wall is observed. This dust component was simplistically modelled using a
uniform density distribution of dust filling one half of the cavity between 160 and 360~AU from the
protostar. In the model coordinate system this dust location within the cavity is $160<R<360$~AU, 
$-150^{\circ} < \Lambda < -45^{\circ}$, where $\Lambda$ is the azimuthal coordinate, defined as a right
handed rotation about the disc axis with an arbitrary zero point. An azimuthal viewing angle of 
$\Lambda = 45^{\circ}$ then qualitatively reproduces Source B so this is the adopted viewing angle 
for the successful model. The optical depth close to the disc axis in the approaching lobe is 
$\tau \sim 0.36$ 
at K band, the dust density being somewhat lower than in the receding lobe. Note that a precise 
3-D fit to the distribution of cavity dust is not possible with 2-D data, so this
aspect of the model is only approximate.

Second, the conical profile of the cavity walls produce a slightly smaller offset of the flux peak at
J band than was observed and the overall shape of the nebula lacks the curvature seen in the data
at J, H and K. As noted in 
section 3, the protostar is slightly offset from the centre of the polarisation pattern (as defined by 
the nulls) due to the inclination of the system (see LR98). The data show an offset along the disc axis 
of $0.40 \pm 0.11$ arcsec from the nulls to the flux peak. The shift in the models is somewhat dependent
on the spatial resolution of the convolution kernel so a conical profile could not be ruled out.
However the conical profile successfully fits the J band data only for a very small region of the 
parameter space in optical depth and vertical density gradient.
Hence, a curvature parameter, $cc$, was introduced to create an approximately parabolic cavity
profile, as described by Eq.(8). A parabola-like profile causes the inner parts of the cavity to 
subtend a larger solid angle to radiation from the protostar than a conical profile; the inner part of 
the cavity then becomes brighter, causing a greater shift of the flux peak at J band. The total
offset in the successful model is therefore composed of 0.15 arcsec from the nulls to the protostar
and a further 0.25 arcsec from the protostar to the flux peak. Note that 
it was also possible to reproduce the J band offset by introducing a knot of dust in the cavity at
the observed location of the flux peak, but this did not reproduce the distortion of the J band 
polarisation pattern which we attribute to dichroic extinction.\\

(8) $((r-R_{cav})/|z|)(R/R_{cav})^{cc}=tan(\theta_c)$\\

{\bf Effects of model parameters}\\

As noted in Section 4, the strong wavelength dependence of the receding lobe in polarised flux and the 
offset of the J band flux peak from the centre of the polarisation pattern are powerful constraints
upon the model. The 'rotating singular isothermal sphere' density distribution in Eq.(7) failed to 
simultaneously reproduce the observed flux and polarised flux ratios of (i) central peak:receding lobe
and (ii) central peak:Source B, for any centrifugal radius in the range $0 < c_{r} \le 1000$~AU.
The failure was worst at J band, where the observed structure could not be even qualitatively
reproduced. The principal difference between the two density distributions of Eq.(6) and Eq.(7)
is in the vertical density gradient. Eq.(6) has a radius-independent vertical density gradient
which appears to fit the data better. This radius-independent flattening might be 
consistent with influence of the magnetic field, causing collapse of the envelope along field lines 
to produce a large disc-like structure.

In summary the effects of the more important parameters are:
\begin{enumerate}
\renewcommand{\theenumi}{(\arabic{enumi})}
\item $v$ (the power law index controlling the vertical density gradient of the envelope (flattening)).
Increasing $v$ causes the receding lobe to become fainter at all wavelengths.

\item $gr$ (grain axis ratio). Increasing the grain axis ratio to make progressively more 
oblate spheroids increases the polarisation of the protostar by increasing the amount of dichroic 
extinction. $gr=1.015$ fits the $\sim 3$ to 4\% polarisation observed with UKIRT at H and K
(see Table 1) when combined with the adopted magnetic field structure (see below). 
$gr=1.03$ is required to fit the larger polarisation (6.8\%) of the protostar measured by Subaru at K band
but it should be noted that the 2 telescopes agree within the measurement errors (see Section 3).
$gr$ would be slightly smaller in the case of a uniform field but could be much larger if the grains 
are not perfectly aligned.

\item $\tau_K$ (envelope optical depth at K band). Values of this parameter in the 
range $2.66< \tau_K < 5.18$ were explored (see section 5.1). The nebulosity in the Subaru data 
appears to have a shallower radial profile than the model: optical depths in the range
3.25 to 4.25 reproduce the flux at radii which increase with decrasing optical depth. 
This suggests that either the radial density gradient of the envelope is less steep than 
$\rho \propto R^{-1.5}$ or there is additional diffuse dust in the cavity, not included in the 
model, which increases the flux at larger radii. Given the apparent asymmetry of the nebulosity
the K band optical depth cannot be more tightly constrained via the radial profile.

\item $\tau_H$ (envelope optical depth at H band). The very prominent receding lobe seen in 
H band polarised flux tightly constrains the H band optical depth to $\tau_H= 5.25 \pm 0.35$, provided 
that system inclinations $i$ consistent with the millimetre observations are employed (see below).
At larger values of $\tau_H$ the receding lobe becomes fainter as the dense disc plane obscures it. At 
smaller values the central peak of the nebula is unresolved in polarised flux, indicating that it is 
dominated by direct flux from the protostar. This contrasts with the data from both telesceopes in which
the central peak is diffuse in polarised flux. At still smaller values, $\tau_H< 4.5$, the receding 
lobe again becomes relatively fainter as the protostar becomes more and more prominent.

\item $\tau_J$ (envelope optical depth at J band). This parameter is constrained by the prominence
and spatial offset of the principal scattering peak observed in total flux, by the faintness of the 
receding lobe, and by the diffuse double peaked structure seen in polarised flux (Figure 4(b)).
Raising the optical depth above the preferred range reduces the prominence of the 'principal' 
scattering peak relative to the more diffuse polarised flux peak, Source B. High optical depth also 
makes the receding 
lobe fainter and can increase the spatial offset. At $\tau_J > 8.4$ Source B becomes much brighter 
than the 'principal' peak. At low optical depth, $\tau_J < 6.65$, the receding lobe becomes too 
bright in diffuse flux and the knot seen at H and K becomes visible at J band, in contrast to 
the data. 

\item $cc$ (cavity curvature). The best fit value is 0.1. 
Increasing $cc$ increases the spatial offset of the J band flux peak
when the empirical envelope density function is used. A positive value of $cc$ (i.e. a more
parabolic cavity) also qualitatively fits the curved morphology of the polarised flux images better
than a conical cavity. 

\item $\theta_c$ (cavity opening angle). This parameter is not tightly constrained since the
values in the range $\theta_c=45^{\circ} \pm 10^{\circ}$ produce acceptable results, and there is some 
degeneracy between $\theta_c$ and the curvature parameter, $cc$.

\item albedo, $\omega$. The albedo of the dust grains was treated as a free parameter (see 5.2.3).
Increasing the albedo increases the prominence of the receding lobe observed in polarised flux, 
relative to the central peak in polarised flux. The ratio of polarised fluxes does not rise linearly 
with albedo however; instead a weaker, approximately square root dependence is seen in the models. 
This behaviour is seen in simulations at all 3 wavelengths, even though the central peak in polarised 
flux is slightly offset from the location of the protostar seen in total flux (the central point source).
For models with inclination $i = 66^{\circ}$, an albedo of $\approx 0.6$ is preferred to precisely
match the prominent receding lobe at H band. However after allowing for the possibility of a slightly
larger system inclination and the fairly weak albedo dependence of the solution, the data are 
consistent with $\omega \gtsimeq 0.4$ at H and $\omega \gtsimeq$ 0.2 at K. The high albedo also helps 
reproduce the morphology of the nebula in polarised flux (see model feature 9 above.) 

\item system inclination, $i$. The inclination derived from the elongation of the inner accretion disc
seen in the millimetre continuum is 62$^{\circ +9}_{-10}$, adopting the most precise available 
determination by Wilner et al.(1997) (see section 5.1). However simulations with 
$i \approx 62^{\circ}$ fail to reproduce the positional shift of the $J$ band flux peak and make the 
receding lobe much too faint in all 3 wavelengths. The minimum inclination which produces a successful 
Monte Carlo fit is $i \approx 66^{\circ}$. Inclinations of $70^{\circ}$ or more can also produce 
successful fits to most of the near infrared data but become less consistent with the millimetre data 
quoted above. The receding lobe also becomes too bright at J band for inclinations much in excess of 
$70^{\circ}$. The preferred inclination is therefore $66^{\circ}<i<71^{\circ}$.
Close et al.(1997) preferred the similar value of $67^{\circ}$ in their Monte Carlo 
simulation, in agreement with the value derived from $^{13}$CO mapping by Hayashi et al.(1993).
\end{enumerate}

\pagebreak
\begin{figure*}
\begin{center}

\begin{picture}(200,440)

\put(0,0){\includegraphics{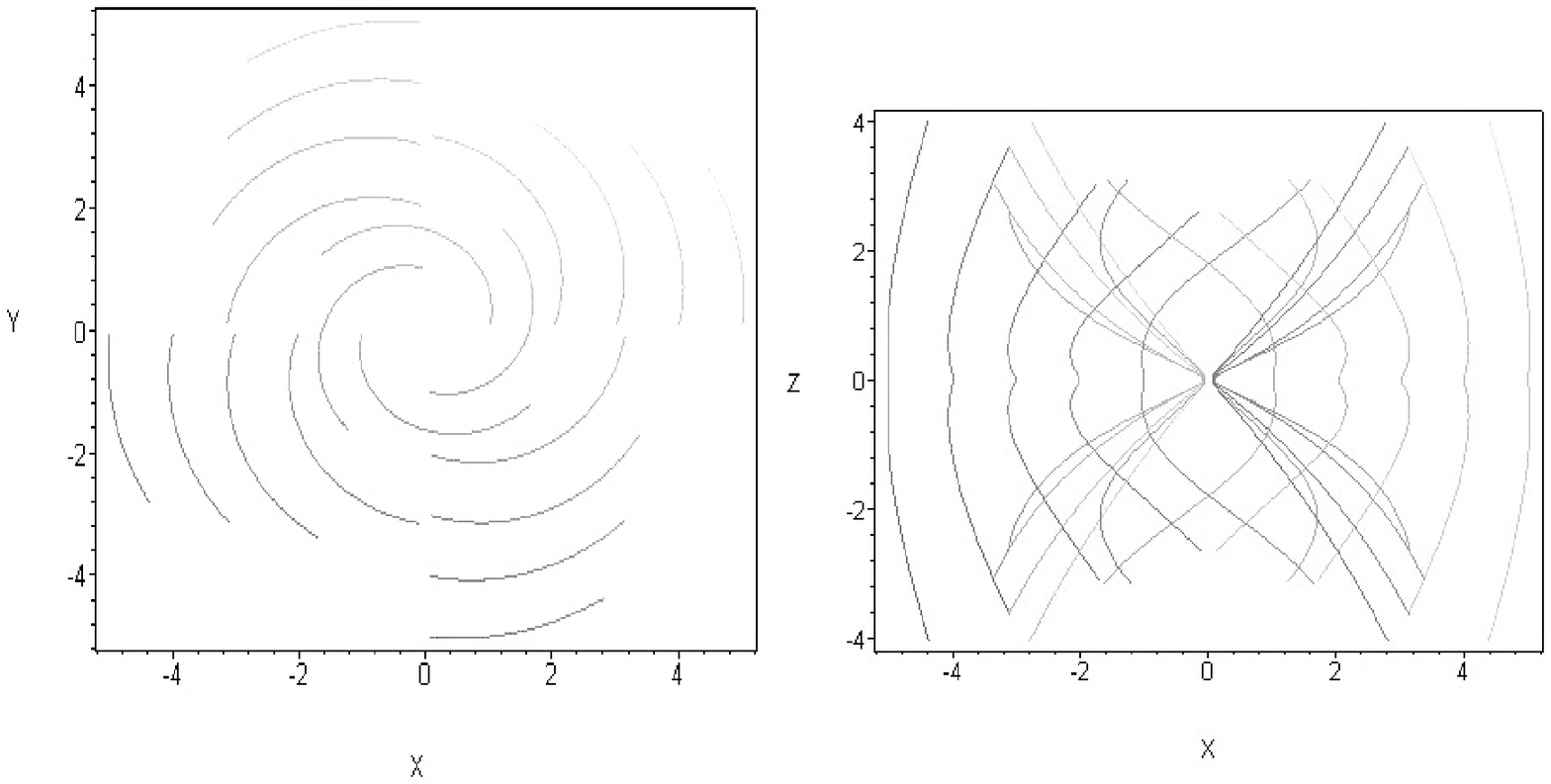}}

\end{picture} 
\end{center}
\small
\vspace{-2cm}
Figure 8. Field structure for parameters $(n=1,m=1,q=1,a=0.4,b0=0.5,b1=1.5,b2=2.0,c=0.1)$,
see Eqs.(9-11). The field lines are highly twisted and also somewhat pinched.
(left) view of the field lines looking down on the disc plane. (right) view
from the disc plane, showing the pinch more clearly.
\end{figure*}

\subsubsection{Structure of the magnetic field}

Polarisation vectors inclined by between 32$^{\circ}$ and 43$^{\circ}$ to the disc axis are observed at the 
location of the protostar in the H and K bands. This can be simply reproduced by 
dichroic extinction by grains aligned with a magnetic field which is globally misaligned with the disc 
axis. This would imply a field orientation of 122$^{\circ}$ to 133$^{\circ}$.
A globally misaligned field cannot be ruled out without spatially resolved circular polarimetry
but an explanation more consistent with star formation theory is that the field is pinched and 
twisted by the collapse of the rotating molecular cloud core from which the presently observed
envelope descended. Such a twisted magnetic field in the circumstellar structure near the central 
star is consistent with that predicted from the millimeter polarization data (Tamura, Hough \&
Hayashi 1995) although their detection is marginal.

We have constructed an empirical formalism for a pinched and twisted field similar to the 
'split monopole' configuration described in recent MHD simulations by Allen, Li \& Shu (2003). 
A detailed fit to the MHD simulations will not be attempted until better quality circular polarisation 
data is obtained. The split monopole indicates the discontinuity in the direction of the field 
lines in the disc plane, which implies the presence of a current sheet. Field lines are assumed to be 
dragged away from an initially axial direction by the rotation of the disc and envelope.

The adopted formalism is:\\

(9) $B_r =  b_1 z/|z| r^m exp(-(r^n + a|z|))$\\

(10) $B_z =  b_0 + b_1/a r^{m-1} exp(-(r^n + a|z|)) (m+1-nr^n)$\\

(11) $B_\phi = b_2 z/|z| (1/(r^q+az^q+c))$\\

where $b_0$, $b_1$, $b_2$, $a$, $c$, $n$, $m$ and $q$ are positive constants 
which parameterise the field structure. One additional parameter
is the unit of length $L_B$ by which the cylindrical polar coordinates $r$ and $z$
are normalised. $B_\phi$ is the field strength in the azimuthal direction perpendicular
to the $\bf r$ and $\bf z$ vectors. This formalism was also used in Lucas et al.(2003).

At large radii ($R \gg L_B$) the field becomes axial ($B_z \rightarrow b_0$, $B_r \rightarrow 0$, 
$B_\phi \rightarrow 0$). The signs of $B_r$ and $B_\phi$ reverse in the plane of the accretion 
disc (z=0) where the pinch and twist is strongest. The form of $B_z$ relates to $B_r$ in a manner 
constrained by the divergence equation. By contrast $B_{\phi}$, the twist component, is independent 
of the other components since the twist is axisymmetric and therefore does not contribute to the 
divergence equation. 

Figure 8 depicts an example of a field structure which successfully reproduces both the 
43$^{\circ}$ inclination of the polarisation of the protostar seen in the UKIRT data and 
the observed distortion of the centrosymmetric pattern, with the spatial scale in units of $L_B$.

The most important parameters in successful field solutions are the coefficients $b0, b1$ and $b2$.
and the scale length $L_B$. The successful model fit in Figure 7 has a cavity radius, R$_{cav}=25$~AU 
in the disc plane, though this parameter is not tightly constrained. Exploration of the magnetic field 
parameter space for this adopted cavity size indicates that for parameters 
($b0, b1, b2=0.5, 1.5, 2.0$, a scale length $L_B \gtsimeq 15$~AU is required to reproduce the 
32$^{\circ}$ to $43^{\circ}$ inclination of the polarisation of the protostar, this being only weakly 
dependent on the power law indices $n$ and $m$. This suggests that while the twist of the field must 
obviously extend into 
the dense inner part of the circumstellar envelope if it is to have a large effect on the dichroic 
extinction, it need not fill most of the volume of the 1300~AU structure. The polarisation gradient 
across the sytem axis observed at J band (see Figure 6) extends to distances of more than 100 AU 
from the protostar, which may indicate that the twist extends to that scale. However the polarisation
gradient is likely to be influenced at some level by the asymmetric distribution of dust in the 
approaching lobe so this might be an over-interpretation of the data.

At $L_B<15$~AU the field is axial over almost the entire envelope volume, causing the 
inclination of the point source's polarisation vectors to the disc axis to become smaller. The 
inclination tends to zero for $L_B \ll R_{cav}$.

\normalsize

In Figure 9(a-c) we illustrate how circular polarimetry would constrain the parameter space. Linear 
polarimetry is sensitive to the field direction only in along lines of sight where the optical
depth is large enough to produce significant dichroic extinction. By contrast circular polarisation
is produced by dichroic scattering (eg. Gledhill \& McCall 2000; Lucas 2003, Whitney \& Wolff 2002)
and is therefore sensitive to the field direction everywhere in the reflection nebula, provided that
there is some grain alignment. Dichroic extinction also influences the circular polarisation state
but this complication will not be significant in the optically thin outer parts of the envelope.




\normalsize
Figure 9 illustrates how circular polarimetry has the potential to reveal the scale length of
any twist in the magnetic field. Circular polarimetry of HL Tau has been attempted on two occasions
using the IRPOL-2 polarimeter at UKIRT. The first attempt during commissioning of the instrument
(Takami et al. 1999) indicated that the HL Tau nebula has circular polarisation $\ltsimeq 1\%$
in the bright central region, though the short exposure time and limited spatial resolution 
($\sim 1$~arcsec) would have inhibited detection of higher circular polarisaton further out in 
the nebula. A second attempt made in 1999 as part of a larger survey (Beckford et al., in prep) 
suggested that there may be $\sim 2\%$ polarisation in the outflow cavity but the data quality was very 
poor so this detection is not secure. A further attempt will be made in better observing conditions.

\hspace{-3cm} \psfig{file=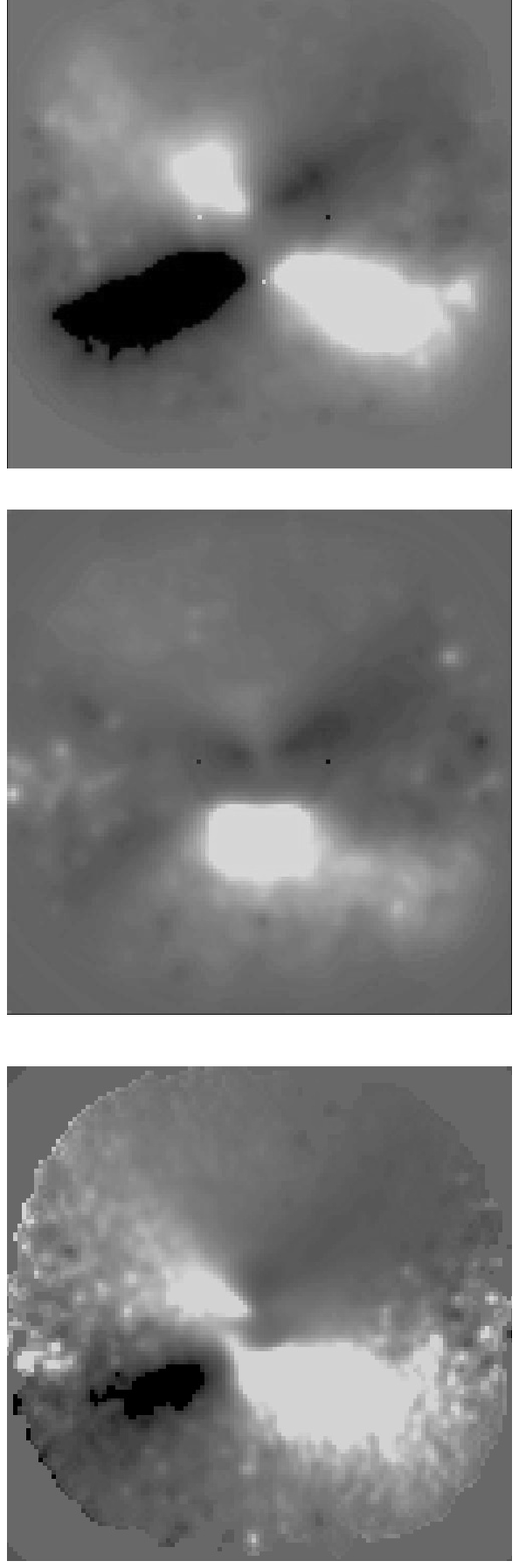,clip=,width=16cm}
\vspace{-7cm}

Figure 9. Model circular polarisation maps for HL Tau (Stokes V/I) at K band in 0.4 arcsec seeing. The 
greyscale ranges from 2\% LCP (white) to 2\% RCP (black), and the models assume $i=66^{\circ}$ and the 
same physical structure as the successful model described in section 5.2.1. (a) map for an axial field, 
showing the commonly observed 4-fold or quadrupolar pattern. (b) map for a helical field pitched at 
45$^{\circ}$ to the disc axis. (c) map for the pinched and twisted field shown in Figure 8 with 
$L_B=100$~AU.

\subsubsection{Nature of the dust grains}

The dust albedo, the near infrared extinction law and the grain axis ratio were treated as free 
parameters for the purpose 
of fitting a successful model to the data. The other optical parameters are (a) the phase function 
and (b) the $4 \times 4$ Stokes matrix and $4 \times 4$ extinction matrix which describe the 
modification of the Stokes vector by scattering and extinction respectively. For these other 
parameters a mixture of ralatively small silicate and amorphous carbon grains was used, which 
produces high linear polarisation at scattering angles near 90$^{\circ}$ and has a relatively 
iostropic phase function. However any mixture of small, almost spherical grains would also have the 
same properties. Small grains in this context means that the scattering cross section of the grain 
mixture is dominated by grains with size parameter $x<1$, where $x= 2 \pi a/ \lambda$, for grain 
equivalent surface area radius $a$, and wavelength $\lambda$.

Having successfully fitted the data we now examine the implications of the solution for the
grain mixture. There are 3 constraints in the successful solution. (1) High linear polarisation
is measured in the outer nebulosity in all 3 filters (up to 75\% at J band, measuring in $3 \times 3$ 
pixel apertures). Similarly high polarisation has been observed in many YSOs in Taurus, eg. 
Whitney et al.(1997); LR98, indicating that small grains dominate the scattering cross section 
in the envelope. To observe such a high polarisation implies that maximum polarisation for single
scattering near $90^{\circ}$ at J band ($P_{Jmax}$) must be at least 75\%, and probably higher. We 
conservatively adopt $P_{Jmax}> 75\%$ as the constraint. 
(2) The extinction law has a ratio of J to K band opacity 
$\kappa(J/K) = 2.1 \pm 0.31$ (see optical depths in Table 2). This is a shallower 
extinction law than that observed in the interstellar medium (see below). The H band extinction is
fully consistent with this result: $\kappa(J/H) = 1.5 \pm 0.14$ and has smaller uncertainty 
than the K band value. A shallow extinction
is most easily reproduced by an absorptive grain mixture which includes some large grains.
(3) The best fit albedo is fairly high: $\omega \gtsimeq 0.4$ at H and $\omega \gtsimeq 0.2$ at K. 
This albedo constraint acts in the opposite sense to the linear polarisation: the size parameter of the 
grains which dominate the extinction cross section cannot be far below unity since the albedo becomes 
negligible in the small grain limit, unless the grains are perfectly dielectric (zero absorptivity).

Comparison of the output of the Mie scattering code 'bhmie.f of Bohren \& Huffman (1989) and the 
ampld.f code for non-spherical particles demonstrates that the slight oblateness implied by our
Monte Carlo fit has a negligible effect on the 3 constraints listed above. We therefore
used the bhmie.f code for uniform spheres and the bhcoat.f code for mantled spheres (also from
Bohren \& Huffman) to search for solutions consistent with the constraints.\\

{\bf Note on extinction law}

The near infrared extinction law implied by $\kappa(J/K) = 2.1 \pm 0.31$ is similar to that
derived by LR98 ($\kappa(J/K) = 1.8 \pm 0.3$), which was principally determined by modelling
of two other low mass YSOs in Taurus, though with qualitative support from additional sources.
Recent imaging at very high spatial resolution (Padgett et 
al.(1999)) has shown that our interpretation of one of these two YSOs, IRAS 04248+2612, was in error. 
The apparently grey central scattering peak was revealed by Padgett et al. as direct light 
from a binary protostar, shining through a region of low extinction in the surrounding envelope.
In the case of HL Tau it is unlikely that unexpected spatial structure in the foreground affects the 
derived extinction law. The $\kappa(J/K) = 2.1$ value is largely based on the ratio of polarised flux 
in the receding lobe to that of the central flux peak. The receding lobe is quite prominent at K 
band, most prominent at H band and very faint at J band, a wavelength dependence that cannot be 
explained by an unexpectedly high or low extinction toward the central source or the receding lobe.

Near infrared extinction in the interstellar medium is observed to be largely independent of the 
reddening parameter, $R_V$, almost certainly because interstellar grains are substantially smaller 
than 1~$\mu$m in radius. We note that the opacity ratio for the interstellar medium is variously 
given as $\kappa(J/K) = 2.52$ (Rieke \& Lebofski 1985); $\kappa(J/K) = 2.48$ (Cardelli, Clayton \& 
Mathis 1989, using a $\lambda^{-1.61}$ law); $\kappa(J/K) = 2.69$ (Tokunaga 2000, using a 
$\lambda^{-1.75}$ law); and $\kappa(J/K) = 2.89$ (Whittet 1992).
By contrast previous Monte Carlo modelling by several authors has used opacities based on theoretical 
interstellar dust models (eg. Mathis, Rumpl \& Nordsieck 1977) which are broadly consistent with
observations across many wavebands but are not particularly successful 
in the near infrared. The corresponding opacity ratio of $\kappa(J/K) \approx 3.25$, which we 
assumed to be correct in LR98, is somewhat larger than the observed values. The difference between 
the interstellar extinction law and the Taurus extinction law is therefore less than earlier 
supposed, but is still likely to be real.

{\bf Grain mixture solutions}

Grain mixtures with an $N(a) \propto a^{-3.5}$ power law and refractive index $n=n^\prime + im^\prime$ 
were explored with the bhmie.f code and it was found that only mixtures with (a) $m^\prime \sim 0.1$; 
(b) $n^\prime$ not far in excess of unity; and (c) a size distribution extending to slightly in excess
of 1~$\mu$m can satisfy the constraints. This region of parameter space is successful because of the
Rayleigh-Gans-like optical properties of large grains with a refractive index near unity.
The polarisation produced by scattering in the Rayleigh-Gans limit is the same as that produced
by Rayleigh scattering (see van der Hulst 1957). The large particles used in successful mixtures 
are not truly in the Rayleigh-Gans limit ($2x (n^\prime-1)\ll 1$) but they exhibit similar polarisation
properties for sizes up to ($x (n^\prime-1) \approx 1$), in just the same way as spheres with a
larger refractive index have Rayleigh-like polarisation for sizes up to $x n^\prime \approx 1$.
Hence large grains with $n^\prime \approx 1$ can have the required combination of small grain
characteristics (high polarisation) and large grain characterisitics (shallow extinction law
and non-negligible albedo). 

\begin{figure*}
\begin{center}
\begin{picture}(200,240)

\put(0,0){\includegraphics{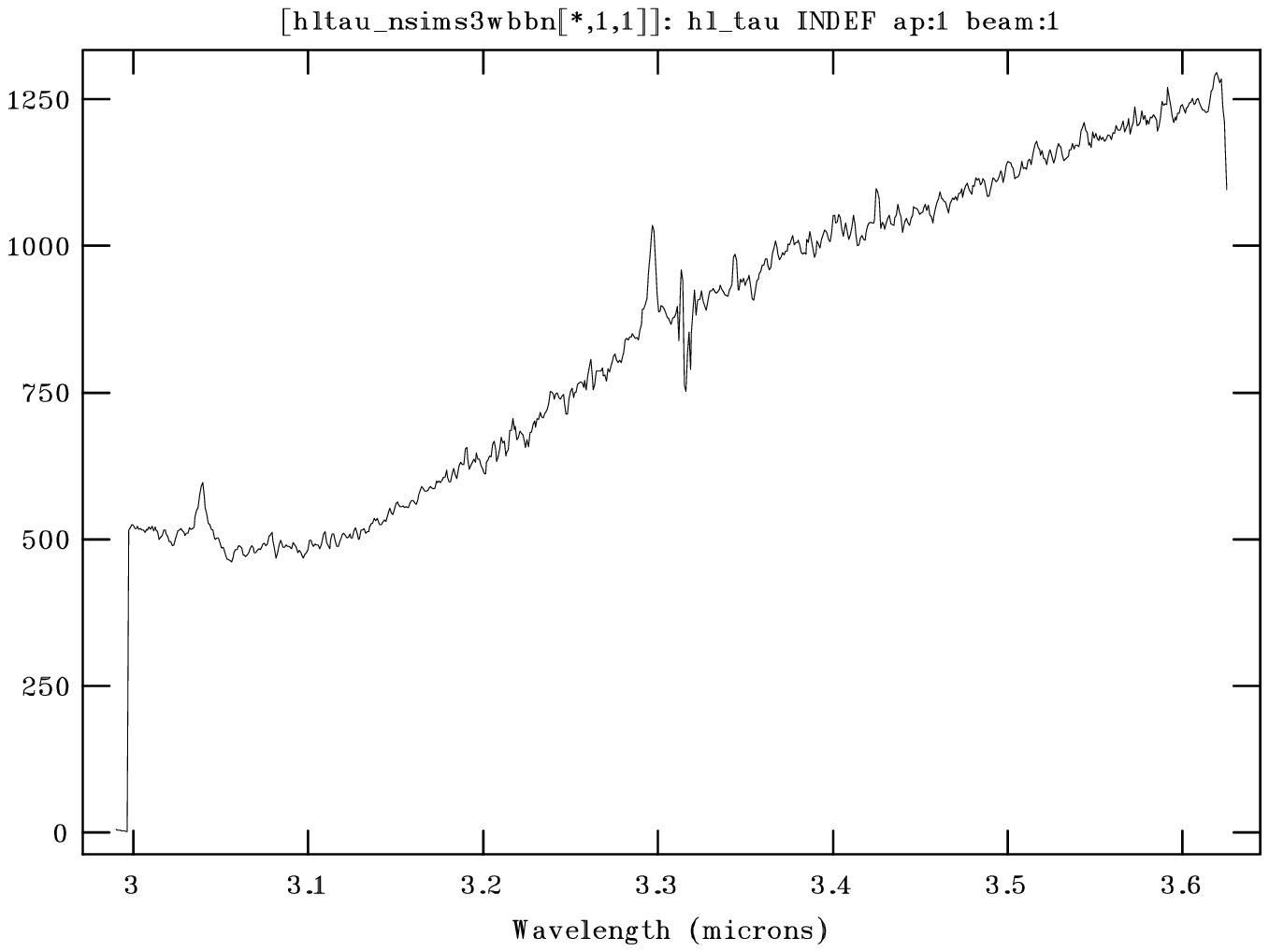}}

\put(-40,215){Pf$\delta$}
\put(-108,215){Pf$\epsilon$}
\put(-34,210){\line(0,-1){5}}
\put(-103,210){\line(0,-1){5}}
\end{picture} 
\end{center}

\small
\vspace{-3cm}
\twocolumn
Figure 10. 3-$\mu$m spectrum of HL Tau. The water ice absorption band centred at 3.05~$\mu$m
is quite strong toward this source, providing qualitative support for a model in which water ice 
dominates the near infrared optical properties of the dust grains. The full extent and depth of the 
ice absorption is difficult to determine since the continuum rises to longer wavelengths, probably
because of continuum emission by hot dust in the inner parts of the accretion disc. Two hydrogen
Pfund series emission lines are marked.
\end{figure*}

\normalsize
A possibly consistent model is that of Pollack 
(1994) in which small silicate grains are surrounded by thick mantles of water ice, intermingled with
more absorptive susbstances. Previous investigation of this model by LR98, using
similar observational constraints and the 'dirty ice' optical constants of Preibisch et al.(1993),
was not very encouraging. However, the computer code 'refice.f' of Warren (1983) provides a somewhat 
smaller value of $n^\prime$ for water ice at 10 to 40~K and the extinction law quoted here is slightly 
less shallow than
previously. These two factors make it possible to manufacture grain models consistent with the data. 
The strong water ice absorption feature observed at 3.05~$\mu$m in HL Tau also lends qualitative 
support to this model (see Figure 10). Water ice is not usually seen in T Tauri stars but is often
observed toward more embedded objects such as HL Tau, see Cohen (1975); Kobayashi et al.(1999); 
Gurtler et al.(1999). Our spectrum is appears consistent with these previous lower resolution data. 
The hydrogen Pfund series lines may prove useful for quantifying the level
of ultraviolet flux in the system, since hydrogen line fluxes at shorter wavelengths are more 
influenced by extinction and scattering. We therefore provide their equivalent widths: 
$0.727 \pm 0.030$~nm for Pf$\epsilon$ and $0.798 \pm 0.050$~nm for Pf$\delta$.

The calculated refractive index for pure ice is $n=1.297+1.3\times 10^{-5}i$ at J band, 
$n=1.288+2.4\times 10^{-4}i$ at H band and $n=1.262+2.6\times 10^{-4}i$ at K band.

In order of increasing sophistication 3 types of model were investigated, the first two 
of which were successful:

\begin{enumerate}
\item A 1 component model: dirty ice grains with $n^\prime$ as quoted above and $m^\prime=0.10$
A size distribution with a $N(a) \propto a^{-3.5}$ in the range $0.01~\mu m < a < 1.5\mu m$ 
and $N(a)=0$ otherwise has the following properties. Maximum J band polariastion, $P_{Jmax} = 75\%$; 
H band albedo, $\omega_H = 0.40$ and opacity ratio $\kappa(J/K)=2.01$.

\item A 2 component model: silicate cores with dirty ice mantles. A core radius of 0.07~$\mu$m is 
adopted, following the interstellar dust model of Li \& Greenberg (1997) and the
refractive indices from Draine (1985) are used. $m^\prime=0.11$ was used for the ice mantles,
with $n^\prime$ as above. A size distribution $N(a) \propto a^{-3.5}$ in the range 
$0.07~\mu m < a < 1.2\mu m$, where $a$ is the total grain radius and only the mantle has a range of 
thicknesses, yields the following properties: $P_{Jmax} = 80\%$; $\omega_H = 0.44$; $\kappa(J/K)=2.14$.

\item A 3 component model: silicate cores with dirty ice mantles as above and separate small 
amorphous carbon grains. Carbon, silicate and water ice are the three most abundant chemical 
constituents suggested 
by Pollack (1994). Carbonaceous inclusions would probably also be required to explain the refractive 
index of the dirty ice. However the observation of Unidentified Infrared Bands (UIR bands) in some 
YSOs at several wavelengths, including the 3.3~$\mu$m wavelength associated with the C-H stretch on 
aromatic 
rings, has been widely attributed to very small carbonaceous particles. Such particles were modelled
by Preibisch et al.1993, with size distribution $N(a) \propto a^{-3.5}$ in the range 0.005 to 
0.03~$\mu$m, the upper bound being imposed by various chemical processes (Sorrell 1990).
We added amorphous carbon grains to the mixture such that the overall size distribution 
was $N(a)=Ka^{-3.5}$; $0.005<a<0.03~\mu$m (amomrphous carbon) and $N(a)=Ka^{-3.5}$ 
$0.07~\mu m < a < a_{max}$ (ice mantled silicates), with the same arbitrary constant K being
used in both cases. However we found that such a model cannot satisfy the observational constraints 
for any combination of $a_{max}$ and ice absorptivity $m^\prime=0.10$, largely because the small
absorptive carbon grains reduce the albedo compared with the equivalent model in (ii) above.  This 
suggests that any very small carbon grains must be less numerous than this simple size distribution 
implies. The absence of 3.3~$\mu$m and 3.4~$\mu$m UIR band emission in Figure 10 is consistent with 
this interpretation, though that might be due to insufficient UV flux for excitation of these 
features.

Finally, we note that alternative grain models with a refractive index near unity could 
probably be constructed using porous grains in which a significant fraction of the volume is 
evacuated. A wide variety of grain compositions could then be considered. However the 
consideration of this far more complicated class of model is beyond the scope of this paper.

\end{enumerate}

\section{Conclusions}

High resolution imaging polarimetry of HL Tau with Subaru and UKIRT has produced very similar results
from both telescopes. 3-D Monte Carlo modelling of the data with aligned non-spherical grains
has allowed us to determine the structure of the circumstellar envelope in considerable detail 
and has demonstrated that dust in the outflow cavities is required to explain many of the departures 
from axisymmetry seen in the polarised flux images.

The envelope structure is found to be only slightly flattened, this being tightly constrained by 
the observations of the receding lobe in the J, H and K bands. An envelope with a radial density
profile $\rho \propto R^{-1.5}$ successfully reproduced the data, although slightly shallower profiles
are also possible. By contrast the Terebey, Shu \& Cassen (1984) density solution for a collapsing and 
rotating singular isothermal sphere, which has an $\rho \propto R^{-0.5}$ radial density profile in 
the inner regions and no flattening in the outer regions, did not reproduce the data successfully.

Both lobes of the bipolar outflow cavity contain a substantial optical depth of dust (not just
within the cavity walls), and curved, approximately parabolic, cavity walls produce a better fit to
the data than conical walls. The small inner accretion disc observed at millimetre wavelengths is not 
seen at this spatial resolution (0.4 arcsec). 

The H and K band polarisation maps show that the polarisation of direct flux from the central point
source (the protostar) is inclined by from $32^{\circ}$ to $43^{\circ}$ to the disc axis. This is s
uccessfully modelled with dichroic extinction by aligned grains. At J band the protostar is obscured 
from view in our data and seen only in scattered light so the polarisation position angle near the 
protostar is more centrosymmetric. However, the J band polarisation vectors in the vicinity of the 
protostar show a distortion of the usual centrosymmetric scattering pattern in both magnitude and 
position angle which is consistent with dichroic extinction of the scattered light. 

Assuming the precession axis of the grains is aligned with the local magnetic field
these data can be modelled with a field which is twisted on 
scales from tens to hundreds of AU, or alternatively by a field which is globally 
misaligned with the disc axis. A unique solution to the field 
structure will require sensitive spatially resolved circular polarisation data.

The Monte Carlo models require a near infrared extinction law which is slightly shallower than
that observed in the diffuse interstellar medium. Additional constraints on the grain optical
properties are very high polarisation in the outer nebulosity and a non-negligible albedo. These 
optical properties can be fitted using grains with a small refactive index. A possible
grain model is composed of silicate cores with dirty water ice mantles in which the largest particles 
have radii slightly in excess of 1~$\mu$m. Such particles have Rayleigh-Gans-like 
polarisation behaviour, even though they are too large to be truly in the Rayleigh-Gans scattering 
regime. An alternative grain model with the required small refractive index might be constructed using
porous particles, which will be investigated in a future paper.

\vspace{4mm}
\textbf{Acknowledgements}\\

We thank the staff of UKIRT for undertaking polarimetry and spectroscopy in the Service 
Programme. UKIRT is operated by the Joint Astronomy Centre on behalf of the UK Particle Physics and 
Astronomy Research Council (PPARC). We also thank the Subaru staff for their support. 
We thank Kenny Wood for the excellent suggestion of using forced scattering and we are grateful to Lumilla 
Kolokolova for the suggestion of a porous grain model. Finally we thank the anonymous referee for an 
invaluable report. PWL is supported by PPARC via an Advanced Fellowship at the University of Hertfordshire. 
MT acknowledges support by Grant-in-Aid (12309010) from the Ministry of Education, Culture, Sports, 
Science and Technology.

\vspace{4mm}
{\bf References}\\

Adams~F.~C., Shu~F.~H., Lada~C.~J., 1988, ApJ, 326, 865\\
Allen A., Li Z.Y., Shu F.H., 2003, ApJ vol. 599, (in press)\\
Bacciotti F., Eisl\"{o}ffel J., Ray T.P., 1999, A\&A 350, 917 \\
Bastien P., M\'{e}nard F., 1988, ApJ 326, 334\\
Beckwith~S., Sargent~A.~I., Scoville~N.~Z., Masson~C.~R., 
   Zuckerman~B., Phillips~T.~G., 1986, ApJ, 309, 755\\
Beckwith S.V.W., Koresko C.D., Sargent A.I., Weintraub D.A., 1989, ApJ, 343, 393\\
Beckwith~S.~V.~W., Sargent~A.~I., Chini~R.~S., G\"{u}sten~R., 1990, 
   99, 924\\
Beckwith~S.~V.~W. Birk~C.~C., 1995, ApJ, 449, L59\\
Bohren C.F., Huffman D.R., 1983, Absorption and Scattering of Light by Small Particles, 
   John Wiley \& Sons, New York\\
Cabrit~S., Guilloteau~S., Andr\'e~P., Bertout~C., Montmerle~T.,
   Schuster~K., 1996, A\&A, 305, 527\\
Cardelli J.A., Clayton G.C., Mathis J.S., 1989, ApJ, 345,245\\
Cashwell E.D. Everett C.J., 1959, A practical manual on the Monte Carlo random walk problems 
   (New York: Pergamon)\\
Close~L.~M., Roddier~F., Northcott~M.~J., Roddier~C., Graves~J.~E., 
   1997, ApJ, 478, 766\\
Cohen M., 1975, MNRAS, 173, 279\\
Draine B.T., 1995, ApJS, 57, 587\\
Draine B.T. Weingartner J.C., 1996, ApJ 470, 551\\
Fischer O., Henning Th., Yorke H.W., 1994, A\&A 284, 187\\
Gledhill~T.~M., Scarrott~S.~M., 1989, MNRAS, 236, 139\\
Gledhill T., Chrysostomou A., Hough J.H., 1996, MNRAS,282,1418\\
Gledhill~T.~M., McCall~A., 2000, MNRAS, 314, 123\\
Greene T.P., Lada C.J., 1996, AJ 112, 2184\\
Gurtler J., Schreyer K., Henning Th., Lemke D., Pfau W., 1999, A\&A 346, 205\\
Hayashi~M., Ohashi~N., Miyama~S.~M., 1993, ApJ, 418, L71\\
Hough J.H., Chrysostomou A.C., Bailey J.A., 1994, Experimental Astronomy, 3, 127\\
Kobayashi, N., Nagata T., Tamura M., Takeuchi T., Takami, H., Kobayashi Y., Sato S., 1999, 
    ApJ 517, 256\\
Lazarian, A. 2003, J. Quantitative Spectroscopy \& Radiative Transfer 79-80, 881\\ 
Li A., Greenberg J.M., 1997, A\&A, 323, 566\\
Lucas P.W., Roche, P.F., 1997, MNRAS 286, 895\\
Lucas P.W., Roche, P.F., 1998, MNRAS 299, 699\\
Lucas P.W., Roche P.F., Allard F., \& Hauschildt P., 2001, MNRAS 326, 695\\
Lucas, P.W., 2003, J. Quantitative Spectroscopy \& Radiative Transfer 79-80, 921\\
Luhman K.L., Briceno C., Rieke G.H., Hartmann, Lee, 1998b, ApJ 493, 909\\
Mathis J.S., Rumpl W., Nordsieck K.H., 1977, ApJ 217, 425\\
Men'shchikov~A.~B., Henning~Th., 1997, A\&A, 318, 879\\
Men'shchikov~A.~B., Henning~T., Fischer~O., 1999, ApJ, 519, 257\\
Mishchenko~M.~I., Hovenier~J.~W., Travis~L.~D., 2000, 
    Light scattering by nonspherical particles : theory, measurements, and
    applications, Academic Press, San Diego\\
Mundt~R., Ray~T.~P., B\"uhrke~T., Raga~A.~C., Solf~J., 1990, 
   A\&A, 232, 37\\
Mundy~L.~G., Looney~L.~W., Erickson~W., Grossman~A., Welch~W.~J., 
   Forster~J.~R., Wright~M.~C.~H., Plambeck~R.~L., Lugten~J.,
   Thornton~D.~D., 1996, ApJ, 464, L169\\
Padgett D.L., Brandner W., Stapelfeldt K.R., Strom S.E., Terebey S., Koerner D., 1999, AJ 117,1490\\
Pollack, J.B., Hollenbach, D., Beckwith, S., Simonelli, D.P., Roush, T., Fong, W., 
   1994, ApJ 421, 615\\ 
Preibisch Th., Ossenkopf V., Yorke H.W., Henning Th., 1993, A\&A, 279, 577\\
Rieke~G.~H., Lebofsky~M.~J., 1985, ApJ, 288, 618\\
Rodr\'iguez~L.~F., Cant\'o~J., Torrelles~J.~M., G\'omez~J.~F., 
   Ho~P.~T.~P., 1992, ApJ, 393, L29\\
Rodr\'iguez~L.~F., Cant\'o~J., Torrelles~J.~M., G\'omez~J.~F., 
   Anglada~G., Ho~P.~T.~P., 1994, ApJ, 427, L103\\
Sargent~A.~I., Beckwith~S.~V.~W., 1991, ApJ, 382, L31\\
Shakura~N.~I., Sunyaev~R.~A., 1973, A\&A, 24, 337\\
Sorrell W.H., 1990, MNRAS, 243,570\\
Stapelfeldt K.R., Burrows C.J., Krist J.E., Trauger J.T., Hester J.J., Holtzman J.A.,
   Ballester G.E., Casertano S., Clarke J.T., Crisp D., Evans R.W., Gallagher J.S.III,
   Griffiths R.E., Hoessel J.G., Mould J.R., Scowen P.A., Watson A.M., Westphal J.A. 
   1995, ApJ, 449, 888\\
Takami~M., Gledhill~T., Clark~S., M\'enard~F., Hough~J.~H., 
   1999, Proceedings of Star Formation 1999; ed. T. Nakamoto, 
   Nobeyama Radio Observatory, p. 205-206\\

Tamura M., Gatley I., Joyce R. R., Ueno M., Suto H., Sekiguchi M., 1991, ApJ 378, 611\\
Tamura M., Hough J.H., Hayashi S.S., 1995, ApJ, 448, 346\\
Tamura~M., Suto H., Itoh Y., Ebizuka N., Doi Y., Murakawa K., Hayashi S.S., Oasa Y., 
  Takami H., Kaifu N., 2000, Proc. SPIE, 4008, 1153\\
Tamura, M., Fukagawa, M., Hayashi, M. et al. 2004, in Proc. IAU Symposium 221 "Star Formation
  at High Angular Resolution" (in press)\\ 
Terebey~S., Shu~F.~H., Cassen~P., 1984, ApJ, 286, 529\\
Tokunaga A.T., 2000, in Allen's Astrophysical Quantities; ed. A.Cox, AIP Press, New York\\
Ulrich~R.~K., 1976, ApJ, 210, 377\\
Warren S.G., 1984, Applied Optics, 23, 1206\\
Weintraub~D.~A., Kastner~J.~H., Whitney~B.~A., 1995, ApJ, 452, L141\\
Weitenbeck~A.~J., 1999, AcA, 49, 59\\
Welch W.J., Hartmann L., Helfer T., Brice\~no C. 2000, ApJ, 540, 362\\
Whitney B.A., Hartmann L., 1993, ApJ, 402, 605\\
Whitney B.A., Kenyon S.J., \& Gom\`{e}z M., 1997, ApJ,485,703\\
Whitney~B.~A. \& Wolff~M.~J., 2002, ApJ, 574, 205\\
Whittet D.C.B., 1992, Dust in the Galactic Environment, pub IOP, London\\
Wilner~D.~J., Ho~P.~T.~P., Rodriguez~L.~F., 1996, ApJ, 470, L117\\
Wolf~S., Padgett~D.~L. \& Stapelfeldt~K.~R., 2003, ApJ, 588, 373\\
Wood~K., Reynolds~R.~J., 1999, ApJ, 525, 799\\
Yamashita~T., Hodapp~K.-W., \& Moore~T.~J.~T., 1994, ApJ, 422, L21\\
Yusef-Zadeh~F., Morris~M., White~R.~L., 1984, ApJ, 278, 186\\

\end{document}